\documentclass{WileyMSP-template}

\usepackage{chemformula} 
\usepackage{pdfpages}

\begin{document}

\rhead{\includegraphics[width=2.5cm]{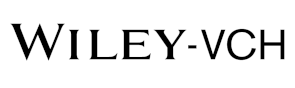}}

\title{Multidimensional Analysis of Excitonic Spectra of Monolayers of Tungsten Disulphide: Towards Computer Aided Identification of Structural and Environmental Perturbations of 2D Materials}

\maketitle


\author{Pavel V. Kolesnichenko}
\author{Qianhui Zhang}
\author{Changxi Zheng}
\author{Michael S. Fuhrer}
\author{Jeffrey A. Davis*}

\begin{affiliations}
Pavel V. Kolesnichenko\\
Optical Sciences Centre, Swinburne University of Technology, Melbourne, Victoria 3122, Australia\\
ARC Centre of Excellence in Future Low-Energy Electronics Technologies, Swinburne University of Technology, Melbourne, Victoria 3122, Australia

Qianhui Zhang\\
Department of Civil Engineering, Monash University, Victoria 3800, Australia

Changxi Zheng\\
School of Physics and Astronomy, Monash University, Victoria 3800, Australia\\
ARC Centre of Excellence in Future Low-Energy Electronics Technologies, Monash University, Victoria 3800, Australia\\
School of Science, Westlake University, Hangzhou 310024, Zhejiang Province, People's Republic of China\\
Institute of Natural Sciences, Westlake Institute for Advanced Study, Hangzhou 310024, Zhejiang Province, People's Republic of China

Michael S. Fuhrer\\
School of Physics and Astronomy, Monash University, Victoria 3800, Australia\\
ARC Centre of Excellence in Future Low-Energy Electronics Technologies, Monash University, Victoria 3800, Australia

Jeffrey A. Davis
Optical Sciences Centre, Swinburne University of Technology, Melbourne, Victoria 3122, Australia\\
ARC Centre of Excellence in Future Low-Energy Electronics Technologies, Swinburne University of Technology, Melbourne, Victoria 3122, Australia\\
Email Address:jdavis@swin.edu.au

\end{affiliations}



\begin{abstract}
Despite 2D materials holding great promise for a broad range of applications, the  proliferation of devices and their fulfillment of real-life demands are still far from being realized. Experimentally obtainable samples commonly experience a wide range of perturbations (ripples and wrinkles, point and line defects, grain boundaries, strain field, doping, water intercalation, oxidation, edge reconstructions) significantly deviating the properties from idealistic models. These perturbations, in general, can be entangled or occur in groups with each group forming a complex perturbation making the interpretations of observable physical properties and the disentanglement of simultaneously acting effects a highly non-trivial task even for an experienced researcher. Here we generalise statistical correlation analysis of excitonic spectra of monolayer WS$_2$, acquired by hyperspectral absorption and photoluminescence imaging, to a multidimensional case, and examine multidimensional correlations via unsupervised machine learning algorithms. Using principle component analysis we are able to identify 4 dominant components that are correlated with tensile strain, disorder induced by adsorption or intercalation of environmental molecules, multi-layer regions and charge doping, respectively.
This approach has the potential to determine the local environment of WS$_2$ monolayers or other 2D materials from simple optical measurements, and  paves the way towards advanced, machine-aided, characterisation of monolayer matter.
\end{abstract}

\section{Introduction}

Since the realization of exfoliation of a single layer of graphite (graphene) and confirmation of its extraordinary physical properties \cite{Novoselov2004}, a wave of efforts aiming at synthesizing other two-dimensional (2D) materials has naturally emerged. A broad spectrum of experimentally obtained ultra-thin materials covering metals \cite{Courty2007,Ma2018}, semimetals \cite{Kim2015}, semiconductors \cite{Splendiani2010}, insulators \cite{Zhang2017}, topological insulators \cite{Kou2017}, superconductors \cite{Menard2017,Menard2019} and ferromagnets \cite{Huang2017} has been already reported with many others having been theoretically predicted \cite{Garcia2011,Andriotis2016,Ding2017,Olsen2019}. This has opened an avenue to material engineering in the form of van der Waals heterostructures giving rise to novel potential devices such as single-molecule and DNA sensors \cite{Arjmandi-Tash2016,Qiu2017}, photodiodes \cite{Furchi2014,Lee2014}, transistors \cite{Georgiou2012}, memory cells \cite{Bertolazzi2013,Choi2013}, batteries \cite{Pomerantseva2017,Zhang2018}, magnetic field sensors \cite{Jimenez2019}, and spintronic logic gates \cite{Kawakami2015,Han2016}.

Despite monolayers holding great promise for a broad range of applications, the research around 2D materials suggests that proliferation of the potential devices and their fulfillment of real-life demands are still far from realization. In contrast to theoretical descriptions of the physical properties of various 2D materials, experimentally obtainable samples commonly experience a wide range of perturbations significantly deviating the properties from idealistic models, and thereby affecting the performance of the devices. Amongst these perturbations are the presence of ripples and wrinkles \cite{Brivio2011,Tapaszto2012,Zhu2012}, point and line defects \cite{Meyer2008,Komsa2013}, grain boundaries \cite{Zande2013}, strain fields \cite{Hsu2017}, doping \cite{Liu2016,Kang2017,Borys2017,Kolesnichenko2019}; water intercalation \cite{Zheng2015}; oxidation \cite{Zhang2013,Hu2019,Kotsakidis2019}, and edge reconstructions \cite{Zhou2013}. These perturbations, in general, can be entangled or occur in groups, forming complex perturbations. This, in turn, makes the interpretations of observable physical properties and the disentanglement of simultaneously acting effects a highly non-trivial task even for an experienced researcher, and advanced characterisation methods are often desirable.

Due to the monolayer nature of 2D materials, their optical signatures are highly sensitive to fluctuations in structure and the local environment. This sensitivity results in non-trivial spatial variations, which are hard to analyse manually, and indicates that unsupervised machine learning algorithms applied to multimodal optical imaging data may aid attempts to identify the fluctuations in structure and the local environment distributed across monolayers. 

Here we consider a semiconducting monolayer of tungsten disulphide (WS$_2$) grown via chemical vapour deposition (CVD) on a sapphire substrate. Optical properties of WS$_2$ monolayers are dominated by excitonic effects manifested as intense signatures in their absorption and emission spectra \cite{Gutierrez2012}. We apply absorption and photoluminescence (PL) hyperspectral imaging to gather data on the spatial variations of the excitonic properties, which arise from the various perturbations. The spectra are fully parameterised, leading to a multi-dimensional parametric phase-space (hypercube) where a single data point represents the set of values corresponding to all parameters at a given spatial location on the monolayer sample. This  then allows us to apply principal component analysis \cite{Pearson1901,Hotelling1933,Lever2017} (PCA) to identify the parameters that vary together and ideally combine to quantify specific perturbations and how they vary across the monolayer flake. A projection of the multi-dimensional data-cloud onto  a 2D plane with axes given by the two most significant principle components preserves the maximum variance in the data. By using unsupervised K-means clustering \cite{Hugo1957,macqueen1967,Lloyd1982} of the data-points in this PCA-plane, regions of the sample with similar properties can be identified and provide further insight into how the perturbations combine and vary across the monolayer sample.

\section{Results and discussion}

\begin{figure}[h!]
\centering
\includegraphics[width=\linewidth,height=\textheight,keepaspectratio]{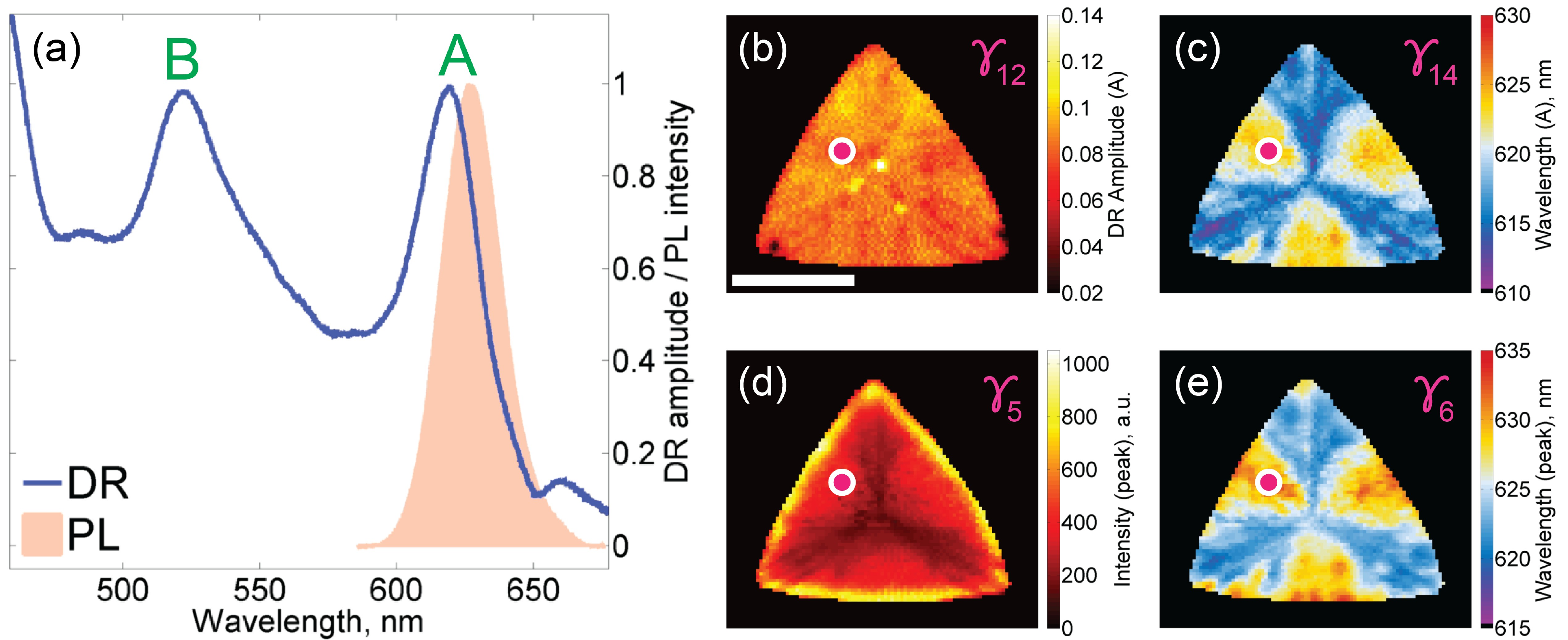}
\caption{(a) Integrated absorption and PL emission spectra of WS$_2$ monolayer. (b,c) DR peak amplitude and wavelength corresponding to A-exciton transition, and (d,e) PL emission peak intensity and wavelength spatial patterns. The length of the scalebar in (b) corresponds to 10 $\mu$m.}
\label{fig:Fig1_PL_DR_maps_typical_spectra}
\end{figure}

Typical absorption and emission spectra of WS$_2$ monolayers are shown in Figure~\ref{fig:Fig1_PL_DR_maps_typical_spectra}a. Absorption spectra are approximated here by differential reflectance \cite{McIntyre1971,Chernikov2015,Borys2017} and feature two distinct peaks corresponding to spin-orbit split A- and B-exciton transitions occurring at K symmetry points in the first Brillouin zone \cite{Liu2015}. Red-shifted PL emission is evident as an asymmetric peak formed as a result of recombination of excitons and trions \cite{Christopher2017}. Figure~\ref{fig:Fig1_PL_DR_maps_typical_spectra}b,c shows the spatially-resolved peak absorption amplitude and wavelength corresponding to the A-exciton transition, revealing trigonally-symmetric variations. Similar trends are observed in the spatial maps of PL emission (Figure~\ref{fig:Fig1_PL_DR_maps_typical_spectra}d,e): the absorption and emission are blue-shifted in the regions spanning from the center of the flake towards its apexes. This type of behaviour has been attributed previously to elevated $n$-doping levels in those areas \cite{Mak2012,Kolesnichenko2019}, and conversely, red-shifted absorption and emission peaks in the adjacent regions have been shown to result from greater tensile strain \cite{Kolesnichenko2019}. While the absorption and emission wavelength maps are somewhat similar, their difference can reveal variations in Stokes shift, which is closely related to charge doping \cite{Borys2017,Kolesnichenko2019}. More obvious differences are observed between the patterns formed by absorption peak amplitudes (Figure~\ref{fig:Fig1_PL_DR_maps_typical_spectra}b) and emission peak intensities (Figure~\ref{fig:Fig1_PL_DR_maps_typical_spectra}d). The most significant difference is that the edges of the triangular monolayer flake can be clearly distinguished in the PL emission intensity map. Along the edges the PL intensity is enhanced as a result of combined effects of water intercalation progressing towards the interior over time \cite{Zheng2015} and oxidation \cite{Hu2019,Kotsakidis2019}. Additionally, three bright spots near the center of the flake can be clearly distinguished in the absorption amplitude map. These bright features are believed to represent multilayer WS$_2$ material formed at the nucleation centers of the monolayer since larger reflectance contrasts have been observed for TMdC multilayers \cite{Frisenda2017}. All these observed differences point to the complementarity of absorption and PL measurements allowing for observations of nonzero correlations between various parameters.

Statistical correlation analysis has been already proven to be a powerful tool in the studies of optoelectronic properties of 2D materials \cite{Lee2012,Bao2015,Hsu2017,Borys2017,Wang2018,Rao2019,Kastl2019}. For example, by correlating spectral shifts of prominent Raman peaks in graphene and graphene/TMdC heterostructures it can be possible to disentangle the effects of doping and strain \cite{Lee2012,Rao2019}. Another route to solve a similar problem for TMdC monolayers used correlations involving the PL Stokes shift \cite{Kolesnichenko2019}. Correlation analyses also facilitated the recognition of physically distinct edges of triangular TMdC flakes as domains hosting large number of point defects \cite{Bao2015,Kastl2019} and the effects of strain on optoelectronic properties of various TMdC monolayers, including direct/indirect nature of the bandgap \cite{Hsu2017}. All these results, however, were based on scatter plots between specifically chosen pairs of parameters missing out other possible correlations, and were not able to recognise the presence of any subtle variations in the data. Here, we generalise statistical correlation analysis to an $N$-dimensional case to acquire more insights into the optoelectronic variations commonly found in 2D materials.

\begin{figure}[h!]
\centering
\includegraphics[width=\linewidth,height=\textheight,keepaspectratio]{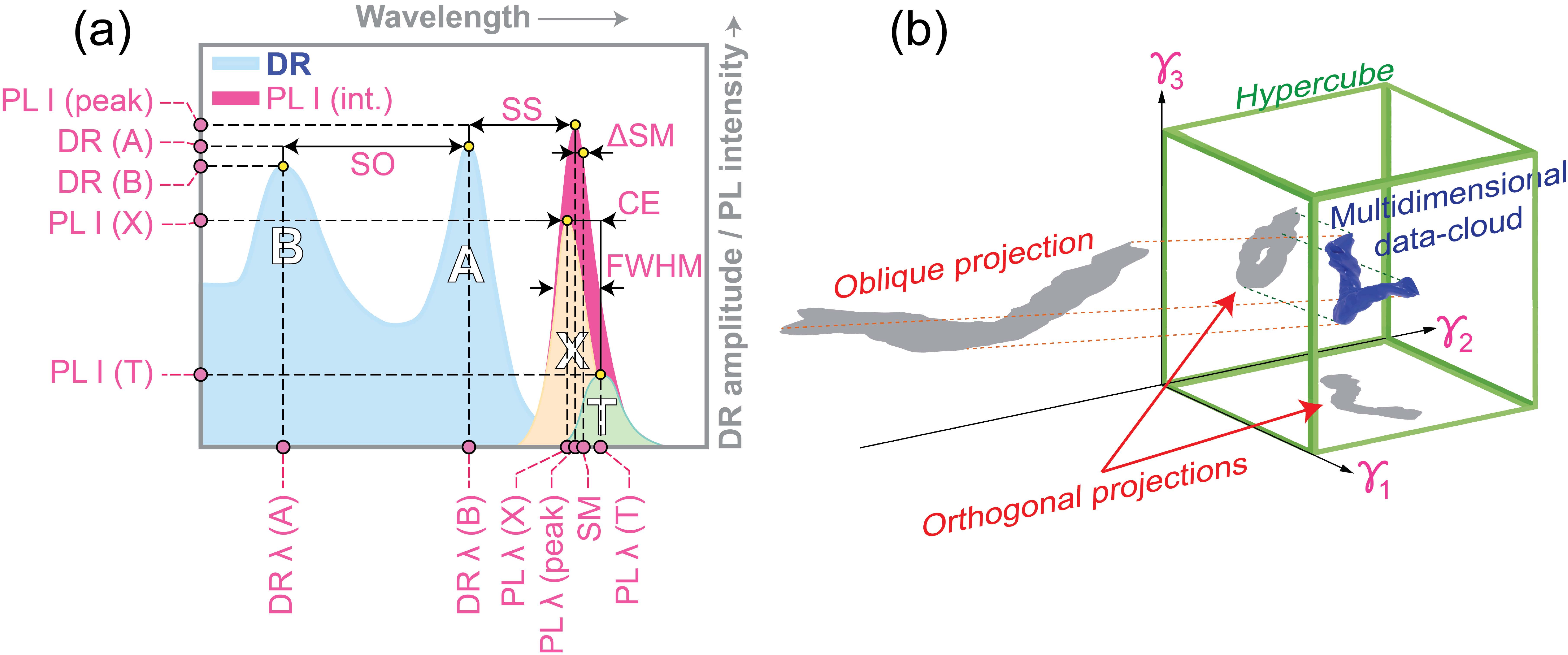}
\caption{(a) Schematic diagram representing absorption (shaded in light blue) and PL emission (shaded in pink) spectra with PL spectrum decomposed into exciton (shaded in yellow) and trion (shaded in green) contributions. All parameters (total 17) used in multidimensional analysis are labeled in pink: PL I (peak) = PL peak intensity; PL I (int.) = PL integrated intensity; SS = Stokes shift; CE = trion charging energy; FWHM = PL full width at half maximum; SM = PL spectral median; PL $\lambda$ (peak) = PL peak wavelength; $\Delta$SM = the difference between the SM and PL $\lambda$ (peak); PL $\lambda$ (X) = exciton emission peak wavelength; PL $\lambda$ (T) = trion emission peak wavelength; PL I (X) = exciton emission peak intensity; PL I (T) = trion emission peak intensity; SO = effective spin-orbit splitting at K symmetry points; DR (A) = differential reflectance peak amplitude of A-exciton; DR (B) = differential reflectance peak amplitude of B-exciton; DR $\lambda$ (A) = differential reflectance peak wavelength of A-exciton; DR $\lambda$ (B) = differential reflectance peak wavelength of B-exciton. (b) Schematic diagram of a multidimensional data-cloud (blue object) within a multidimensional hypercube. Qualitatively different trends (shaded in grey) can be observed depending on the angle of view. Generic parameters $\gamma_1, \gamma_2, \gamma_3, ..., \gamma_N, N=17$, form dimensions (axes) of the hypercube.}
\label{fig:Fig2_Parameters_Hypercube_Projections}
\end{figure}

To make the $N$-dimensional correlation analysis possible we fully parameterise absorption and emission spectra (Figure~\ref{fig:Fig2_Parameters_Hypercube_Projections}a) and use each of the parameters to represent a dimension of an $N$-dimensional parametric space (Figure~\ref{fig:Fig2_Parameters_Hypercube_Projections}b), with $N=17$ in our case. This space is represented by an $N$-dimensional hypercube ($N$-cube) encapsulating an $N$-dimensional data-cloud where each data-point $\vec{\gamma}$ is described by a set of $N$ values (coordinates), i.e. $\vec{\gamma}=\{\gamma_1,\gamma_2,...,\gamma_N\}$. The parameters whose spatial variations are mapped in Figure~\ref{fig:Fig1_PL_DR_maps_typical_spectra}b--e correspond to $\gamma_5,\gamma_6,\gamma_{12},\gamma_{14}$. The definitions of the other parameters and their spatial distributions are given in Supporting Information (Table~1 and Figure~1, respectively). A specific location (pink point in Figure~\ref{fig:Fig1_PL_DR_maps_typical_spectra}b--e) on the monolayer island can, therefore, be assigned a set of $N=17$ numbers corresponding to the values of the 17 parameters chosen to describe the optical properties of the material. We note, that some parameters measure similar quantities (e.g. PL peak intensity and PL integrated intensity), however, there are subtle differences which can be important and so at this stage we include all of them for completeness.

The natural approach to visualization of the geometry of a multi-dimensional  object (data-cloud) is to look at its projections onto 2D planes (Figure~\ref{fig:Fig2_Parameters_Hypercube_Projections}b). Amongst infinite number of possible planes and projecting angles, a particular case of orthogonal projections onto the sides of the $N$-cube is the simplest to realise. It is this particular case that was considered in the previously reported correlation analyses of optoelectronic properties of 2D materials where certain physical trends and clusters have been identified \cite{Lee2012,Bao2015,Hsu2017,Borys2017,Rao2019,Kastl2019,Kolesnichenko2019}. In some instances, oblique projections can provide greater separation of the data, and in some cases correspond to meaningful parameters. For example, charging energy is defined as the difference between the exciton and trion energy, and would correspond to an oblique projection, as detailed in the Supporting Information.

\begin{figure}[h!]
\centering
\includegraphics[width=1\linewidth,height=\textheight,keepaspectratio]{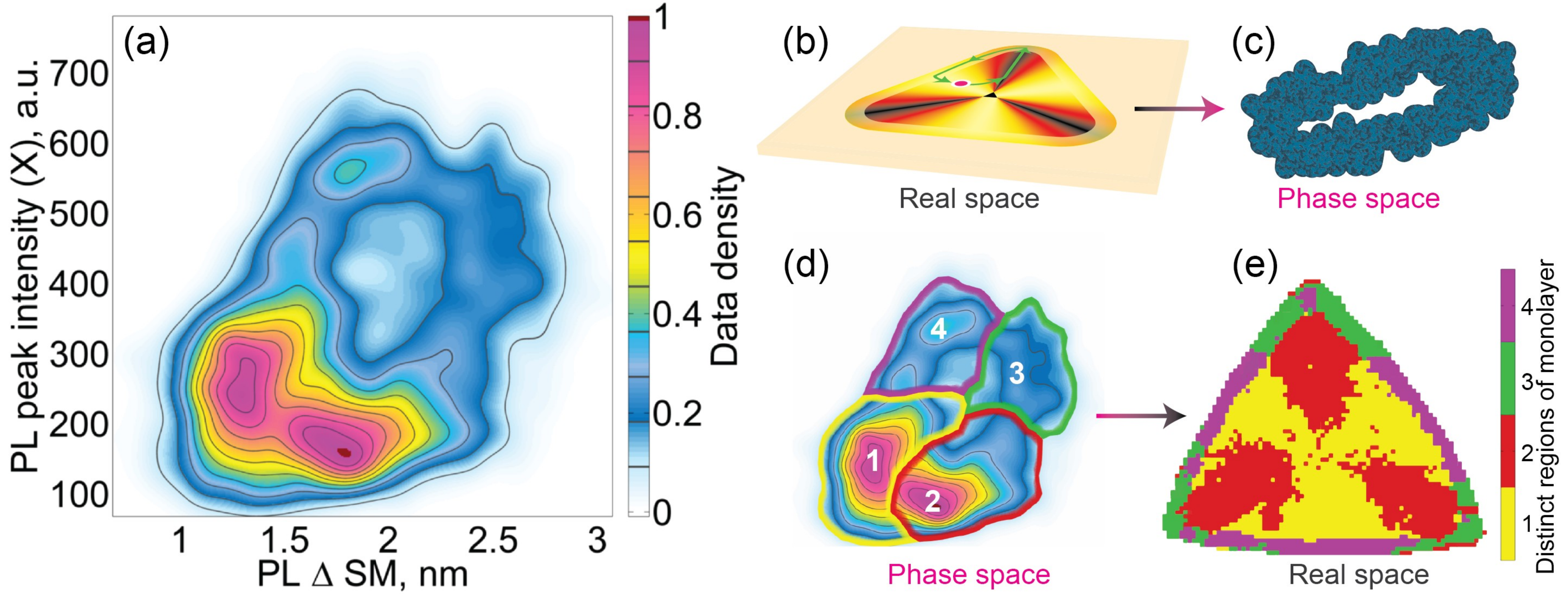}
\caption{(a) 2D orthogonal projection of the data onto the plane formed by the exciton PL peak intensity and PL $\Delta$SM featuring a ``shadow'' of a torus. The data-density is colorcoded, and the levels of contours are marked on the colorbar. (b,c) Schematic diagram demonstrating how a torus (c) can be obtained in the multi-dimensional parametric phase space for the case of WS$_2$ monolayer characterised by the optical spectroscopy in the real space (b).  (d,e) Initial mapping of the data represented in the phase space back into the real space. Four heterogeneous domains are identified: heterogeneous interior (red and yellow) and heterogeneous edge (green and purple).}
\label{fig:Fig3_Experimental_Ring}
\end{figure}

An example of an orthogonal projection is shown in Figure~\ref{fig:Fig3_Experimental_Ring}a. In this case, the data is distributed in a ring, which indicates that the full data-cloud is a torus-isomorphic object (other projections showing torus-shaped data-density distributions are shown in Supporting Information). This topology can then also be related to the spatial variations of material properties assuming they vary smoothly, which is typically the  case. Specifically, the shape of the data-cloud indicates that it is possible to define loops where each point on the loop has distinct spectral properties (Figure~\ref{fig:Fig3_Experimental_Ring}b--c). These loops will be around a specific point or points on the material.

To help visualise this we note that around the data-cloud ring in Figure~\ref{fig:Fig3_Experimental_Ring}a, there appears to be four main clusters of data, as identified in Figure~\ref{fig:Fig3_Experimental_Ring}d. The boundaries are defined as the lines connecting those points of the data density contours that have high negative curvature \cite{Morisawa1957,Pelletier2013} (see Supporting Information for estimation of curvatures). The data-points within each cluster are mapped back into their real-space location in Figure~\ref{fig:Fig3_Experimental_Ring}e, showing that the clusters are indeed related to specific regions on the sample. The points about which the circular paths can be identified, are the points where the four colours (corresponding to the four clusters) meet. 
It is apparent then that by selecting specific orthogonal projections and subsequent clustering analysis, we can gain some insight into how the optical properties (and corresponding structural/environment properties) vary across the sample. However, it is likely that there is further fine structure in these clusters, indeed, it is expected that many of the sample perturbations vary smoothly and continuously.
These perturbations will affect the 17 different parameters in different ways, and in most cases will affect more than one parameter. To identify the parameters that vary together and maintain the maximum variance of the data,
we apply the principal component analysis (PCA) \cite{Pearson1901,Hotelling1933,Lever2017}. 
By identifying the orthogonal principle components (i.e. specific linear combinations of the 17 parameters) that maintain the maximal variance, the ability to resolve fine structure and small variations in the data cloud is enhanced. Furthermore, in identifying the parameters that vary together, this approach has the potential to separate each specific sample perturbation (e.g. strain, doping, molecular adsorption, etc.) and the specific changes to the optical properties induced by each of them. It may then become possible to map the structural and environmental properties across the sample.

The method for PCA has been described previously \cite{Pearson1901,Hotelling1933,Lever2017} and the details of the approach used here are included in the Supporting Information. The result is a new set of axes (the principle components) for the data hypercube. These principle components are defined such that the variances $\Delta_i, i=\{1,...,17\}$ of the data along the principal components (PC$i$) decrease for each successive component (i.e. $\Delta_1\geq\Delta_2\geq...\geq\Delta_{17}$) with $\Delta_1$ corresponding to the maximal variance of the data (distributed along PC1). The corollary of defining the principle components in this way is that it also identifies the measurement parameters that vary together, and distills the variance of the 17-dimensional data-cloud into a hypercube of a lower dimensionality. In the case of the hypercube defined by the parameters of absorption and emission spectra considered here, the PCA approach showed that it is possible to reduce the number of dimensions from 17 (defined by the parameters $\gamma_1,...,\gamma_{17}$) down to 4 (defined by the parameters PC1, ...,PC4) and still preserve as much as 87.9\% of the total variance (see Supporting Information).

Each of the principle components are formed by a linear combination of the measurement parameters. The relative weight of the parameters for each PC is shown in the Supporting information, Table~2. The amplitude of each principle component then varies across the sample and can be mapped spatially in the same way the measurement parameters were mapped in Figure~\ref{fig:Fig1_PL_DR_maps_typical_spectra}b--e. Based on the make-up of each PC, their spatial variations across the WS$_2$ flake, and previously reported understanding of these materials, we are able to correlate each PC with a specific perturbation (or group of perturbations) of the sample structure and/or environment. Specifically, we link PC1 with variations in strain, PC2 with disorder induced by adsorption and/or intercalation of environmental molecules, PC3 with multilayers and PC4 with charge doping.

\begin{figure}[h!]
\centering
\includegraphics[width=1.0\linewidth,height=\textheight,keepaspectratio]{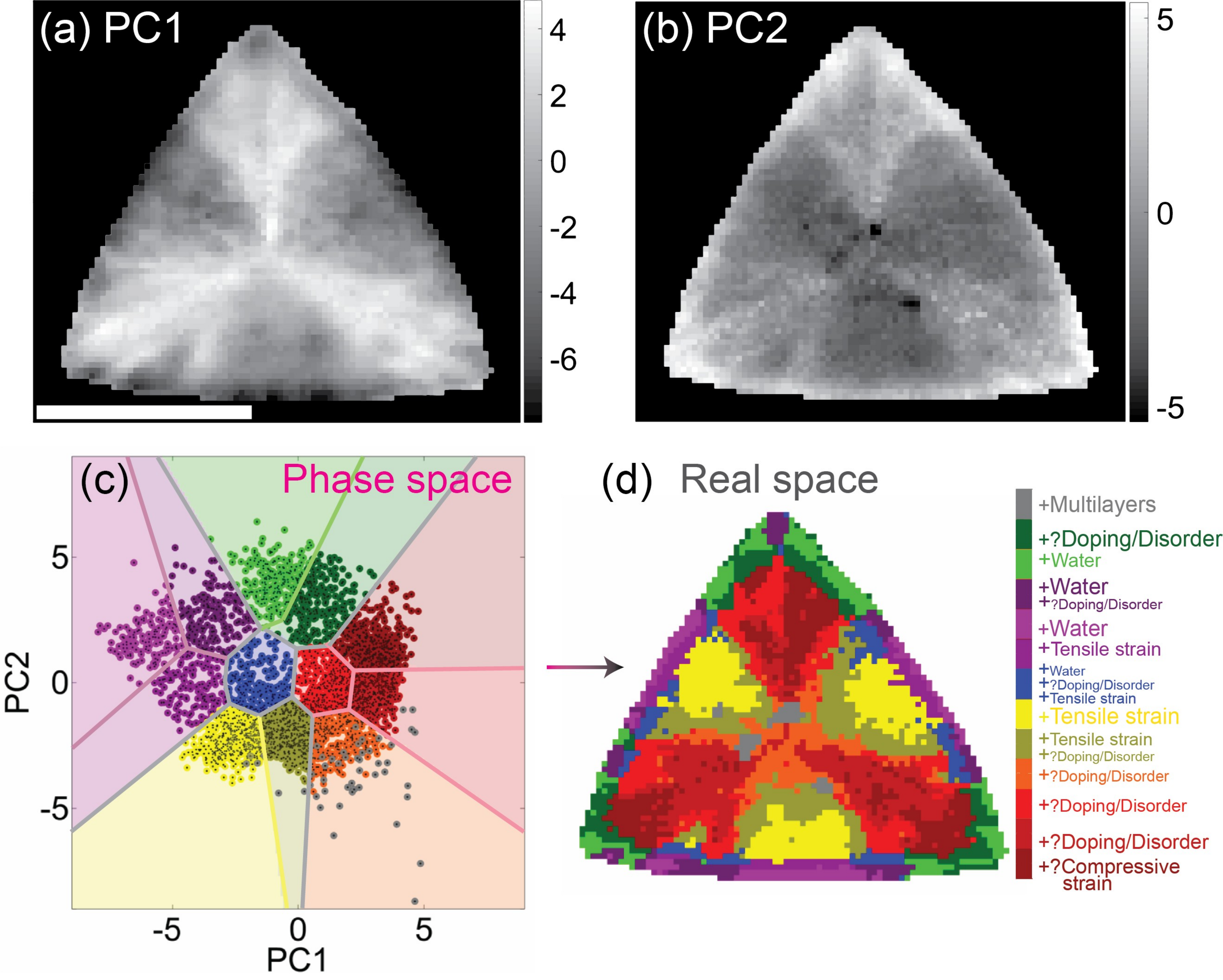}
\caption{Spatial maps of the two most significant components, (a) PC1 and (b) PC2. The scalebar in (a) corresponds to 10~$\mu$m. (c) 2D projection of 17-dimensional data-cloud onto the PCA-plane spanned by the parameters PC1 and PC2. Colored areas are the domains separated by the K-means clustering algorithm featuring 4 main clusters (yellow, green, red and purple) and 12 subclusters (shades of yellow, green, red and purple). Gray data-points correspond to multilayers identified by an anomaly detection method. A blue-colored cluster is centered around 0 in the PCA plane and corresponds to a ``boundary'' between the four main domains. (d) Real-space: all 13 clusters (including multilayers) are mapped back onto the WS$_2$ monolayer flake. Colorbar is labeled in accordance with the previously reported results. A question mark ``?'' in front of a label indicates an unconfirmed and tentative assignment. The size of the labels' font symbolically represents the weight of the corresponding perturbation. 
}
\label{fig:Fig4_PCA_KMeans_labels}
\end{figure}

The attribution of PC1 to strain variation is based on the observation that the dominant contributions to this component are the PL intensity and wavelength parameters, as well as the absorption wavelength for the A-exciton, consistent with previous measurements reporting the effect of strain on optical properties  \cite{Zhang2016,Frisenda2017,Niehues2018}. In addition, the spatial variation of PC1 across the sample (Figure~\ref{fig:Fig4_PCA_KMeans_labels}a) is consistent with previous observations of how the strain varies from the apexes to the middle of the sides in CVD-grown WS$_2$ monolayers  \cite{McCreary2016,McCreary2017,Kolesnichenko2019}. There are other perturbations that can also alter the PL intensities and wavelengths, however, these also affect other parameters that are absent from this principle component (e.g. doping also leads to substantial Stokes shift). 
PC2 (Figure~\ref{fig:Fig4_PCA_KMeans_labels}b) is dominated by variations in the PL FWHM, $\Delta$SM, and charging energy (CE). Alone, $\Delta$SM and CE can be associated with doping density, however, a clearer signature of doping is the Stokes shift  \cite{Mak2012,Borys2017,Kolesnichenko2019}, which doesn't make a significant contribution to this PC. Furthermore, this is a flake that has been exposed to air for some time and previous work has shown that where freshly grown flakes have large amounts of $n$-doping in the region of the apexes, the aged flakes adsorb environmental molecules, which reduce the density of free charges, and increase the FWHM  \cite{Kim2016,Borys2017,Sun2017,Kolesnichenko2019}. The spatial variation of PC2 adds further weight to this assignment, as in addition to the expected variations in the bulk of the 2D flake, it also reveals the edges where water intercalation has occurred  \cite{Zheng2015,Kolesnichenko2019}. 

PC3 is dominated by the DR peak intensity for both A- and B-excitons, and the spatial map shows the most significant variance occurs in small regions near the centre of the flake. This is consistent with previous measurements that have attributed significant increases in DR intensity to increased scattering from multi-layer regions on the sample  \cite{Dhakal2014,Castellanos-Gomez2016}. 
The dominant contribution to PC4 is the PL Stokes shift, which is strongly linked with charge doping  \cite{Mak2012,Borys2017,Kolesnichenko2019}. The relatively low level of data variation in PC4-map is consistent with previous observations of small variations of doping density on aged CVD-grown flakes. Interestingly, the other significant contribution to PC4 is the effective spin-orbit splitting, or in other words the energy difference between the A- and B-excitons in the DR measurements. It was previously speculated that these variations could be due to increased doping \cite{Kolesnichenko2019}, but other possibilities couldn't be ruled out. This observation adds further weight to the case that it is due to variations in doping density.

The ability to self-consistently attribute specific sample perturbations to the four primary principle components that arise naturally from PCA of a single flake is potentially of great value. It points to the possibility of developing a set of well-defined principle components, consisting of linear combinations of spectroscopic parameters, that could be applied to determine the precise structural and environmental perturbations at a specific location in a 2D semiconductor. Further analysis of how these perturbations vary across the flakes could then provide insight into growth mechanisms and causes of the variations from pristine materials. We note, however, that to have a greater level of confidence in the make-up of the significant principle components and their relationship to physical perturbations, analysis of a larger dataset and materials prepared under different conditions with different combinations of perturbations is required.  This will be the subject of future work.

One of the challenges that may be resolved by this type of PCA is the ability to distinguish different contributions, where multiple perturbations occur at the same place. To demonstrate this we project the data onto the plane spanned by the first two principal components (PC1 and PC2) (Figure~\ref{fig:Fig4_PCA_KMeans_labels}c), associated with strain and the adsorption of environmental molecules and/or intercalation of water. This projection shows the maximum spread of the data and reveals some clustering of data points.
In order to acquire more insights into the projected data and correlate the coordinates in this PC1-PC2 plane with the spatial location on the WS$_2$ flake, we applied the K-means-clustering unsupervised learning algorithm \cite{Hugo1957,macqueen1967,Lloyd1982}. This algorithm, for a given input number $K$, tries to classify the data-set into $K$ labeled clusters (see Supporting Information for details). The value of $K$, however, cannot be automatically identified by the algorithm, and, therefore, cluster identification methods are commonly used. Here we used the so called ``elbow'' method \cite{Aldenderfer1984} as one of the most popular methods for identification of the natural number of clusters, if there are any (see Supporting Information). As expected, in the case of the WS$_2$ monolayer considered here, the ``elbow'' method revealed the presence of four prominent clusters in the data-cloud (see Supporting Information) corresponding to the two heterogeneous interior domains and two heterogeneous edge domains within the flake, as identified above in Figure~\ref{fig:Fig3_Experimental_Ring}. However, the method also revealed that there are other natural cluster sets ($K=2$, $K=8$ and $K=12$) present in the data, although they are not as prominent as the set of 4 clusters ($K=4$). With increasing the number of clusters $K$, additional ``shades'' are introduced to the $K=4$ cluster set (Figure~\ref{fig:Fig4_PCA_KMeans_labels}c). We note that we excluded multilayer regions, which are separated from the rest of the data cloud in the PC3 component (gray data-points in Figure~\ref{fig:Fig4_PCA_KMeans_labels}c), from the K-means analysis by treating them as ``anomalies'' (or ``outliers'') 
\cite{Chandola2009}.

We then mapped the data-points in the PCA-plane within each of the clusters back onto the monolayer flake (Figure~\ref{fig:Fig4_PCA_KMeans_labels}d) revealing fine-structure of the four main regions of the sample mentioned above. These four clusters can be clearly grouped in pairs, with the purple and green regions mapping the edges affected by water intercalation, and the red and yellow regions mapping the interior of the flake. This roughly correlates with the higher values for PC2 around the edges, although it can also be seen that the value of PC2 increases when going from purple to green. A similar trend is seen when going from yellow to red in the interior. This indicates that while the main change in going from yellow to red and purple to green is along the PC1 axis, and due to reducing tensile strain, this trend is accompanied by an increase in disorder due to adsorbed molecules, and hence a shift along the PC2 axis. 

We note also that water intercalation does not change appreciably the amount of strain along the edges, as evidenced by the consistent variation along PC1 for both the edges and interior. This suggests that on average the strain field vectors are aligned angularly around the center of the monolayer island so that the radially-propagating water intercalation does not release strain (see also Supporting Information, Section 10).

\section{Conclusions}

These results indicate that the principle component analysis based on the spectral parameters from PL and DR hyperspectral imaging, has the potential to disentangle and quantify different types of perturbations in monolayer materials. This approach is effectively an extension of specific 2D correlation plots that have been used to help understand the variations across a monolayer  flake \cite{Lee2012,Bao2015,Hsu2017,Borys2017,Rao2019,Kastl2019,Kolesnichenko2019}, and which were used here to reveal different regions of the monolayer flake with clearly different combinations of perturbations. 
The principle component analysis, however, is a more systematic and quantitative approach. 

The PCA applied to the data here produced four dominant, orthogonal principle components. By examining the combination of spectral parameters that makes up each of these, and the variation of these PCs across the flake, we were able to assign a specific sample perturbation to each: tensile strain, disorder induced by adsorption/intercalation of environmental molecules, multi-layers, and charge doping. These assignments, the spatial variations and spectral parameters contributing are fully consistent with previous measurements and understanding developed from similar flakes \cite{Dhakal2014,Castellanos-Gomez2016,Zhang2016,McCreary2016,McCreary2017,Frisenda2017,Niehues2018,Kolesnichenko2019}. However, these assignments are not definitive and may not be able to reliably predict the specific perturbations on a different flake. It does, however, point to the possibility of using this approach for this purpose. To achieve this, a larger sample size including multiple flakes with different levels of the different perturbations is needed, and a refinement of the parameters may be necessary to remove the intensity/amplitude parameters, which depend on the measurement system. Subsequent steps could involve a large labeled dataset, and combinations of PCA and cluster analysis to train neural networks and enable real-time identification of spatially-varying perturbation. Regardless, the demonstration here that a self-consistent attribution can be made using PCA on a single flake indicates that this approach is promising, and may allow creation of a tool capable of identifying the perturbations at a given location in a given 2D material simply from PL and DR spectra.

\section{Experimental Section}

\subsection{Sample preparation}

The sample preparation was performed in a similar way as described in Ref.~\cite{Zhang2018a}. Briefly, monolayers of WS$_2$ were grown on sapphire substrate via CVD using WO$_3$ and sulphur precursors. WO$_3$ precursor was placed in the middle of the chamber (high-temperature zone, 860$\degree$C) while the sulphur precursor was placed further upstream (low-temperature zone, 180$\degree$C). The substrate was placed in a direct proximity to the WO$_3$ precursor. Heating the chamber and maintaining the hot environment lasted over $\sim$45 min, after which a cooling process was initiated.

\subsection{Experimental realization}

The PL and DR imaging setups were implemented in the same way as described in Ref.~\cite{Kolesnichenko2019}. In a nutshell, in PL experiments, linearly polarised cw radiation ($\sim$410~nm) was focused on the sample through a 100x objective lens (NA=0.95). The detection scheme was implemented in an epi-fluorescence geometry in a confocal way with the detection spatial resolution reaching $\sim$300~nm. DR measurements were performed using a broadband (400--800~nm) incoherent white light radiation with the detection spatial resolution reaching $\sim$380~nm.

\medskip
\textbf{Supporting Information} \par 
Supporting Information is available from the Wiley Online Library.

\medskip
\textbf{Acknowledgements} \par 
This work was supported by the Australian Research Council Centre of Excellence for Future Low-Energy Electronics Technologies (CE170100039).





\medskip
\bibliographystyle{MSP}
\bibliography{acsnano_paper_PK.bib}



\section*{Supplementary material}

# Supporting Information

# Multidimensional analysis of excitonic spectra of monolayers of tungsten disulphide: Towards computer aided identification of structural and environmental perturbations of 2D materials

Pavel V. Kolesnichenko,*,†,¶ Qianhui Zhang,‡ Changxi Zheng,‡,§ Michael S. Fuhrer,‡,§ and Jeffrey A. Davis*,†,¶

†Centre for Quantum and Optical Science, Swinburne University of Technology, Melbourne, Victoria 3122, Australia

‡Monash University, Melbourne, Victoria 3800, Australia

¶ARC Centre of Excellence in Future Low-Energy Electronics Technologies, Swinburne University of Technology, Melbourne, Victoria 3122, Australia

§ARC Centre of Excellence in Future Low-Energy Electronics Technologies, Monash University, Victoria 3800 Australia

E-mail: pkolesnichenko@swin.edu.au; jdavis@swin.edu.au

# Contents

# S1. Hypercube's dimensions

Table 1: List of parameters $\gamma_i, i = \{1, ..., 17\}$, used as dimensions for a hypercube. "X" stands for exciton; "T" stands for trion; "A" stands for A-exciton; "B" stands for B-exciton; "SM" stands for spectral median.

| Parameter | Meaning | Parameter | Meaning |
|---|---|---|---|
| $\gamma_1$ | PL peak intensity (X)[a] | $\gamma_{10}$ | PL FWHM |
| $\gamma_2$ | PL peak wavelength (X)[a] | $\gamma_{11}$ | Trion charging energy[a] |
| $\gamma_3$ | PL peak intensity (T)[a] | $\gamma_{12}$ | DR peak intensity (A) |
| $\gamma_4$ | PL peak wavelength (T)[a] | $\gamma_{13}$ | DR peak intensity (B) |
| $\gamma_5$ | PL peak intensity | $\gamma_{14}$ | DR peak wavelength (A) |
| $\gamma_6$ | PL peak wavelength | $\gamma_{15}$ | DR peak wavelength (B) |
| $\gamma_7$ | PL integrated intensity | $\gamma_{16}$ | Effective spin-orbit splitting |
| $\gamma_8$ | PL spectral median | $\gamma_{17}$ | PL Stokes shift |
| $\gamma_9$ | PL $\Delta$SM | | |

[a] These parameters were derived from fittings;

Various parameters extracted from hyperspectral absorption and PL emission imaging can be regarded as dimensions of a parametric phase-space represented by a hypercube. In this work 17 different parameters listed in Table 1 were used to construct a 17-dimensional hypercube. Spatial maps corresponding to each of these dimensions are shown in Figure 1 (except for those given in Figure 1b–e in the main text). As seen from Figure 1, a certain spatial location $\vec{\gamma}$ (purple point) can be represented by a set of 17 numbers (i.e. $\vec{\gamma} = \{\gamma_1, \gamma_2, ..., \gamma_{17}\}$) and, therefore, by a single point in the constructed hypercube. It is possible that several spatial locations on the monolayer flake can be represented by a similar set of the chosen 17 parameters resulting in the points located close to each other in the parametric phase-space forming clusters.

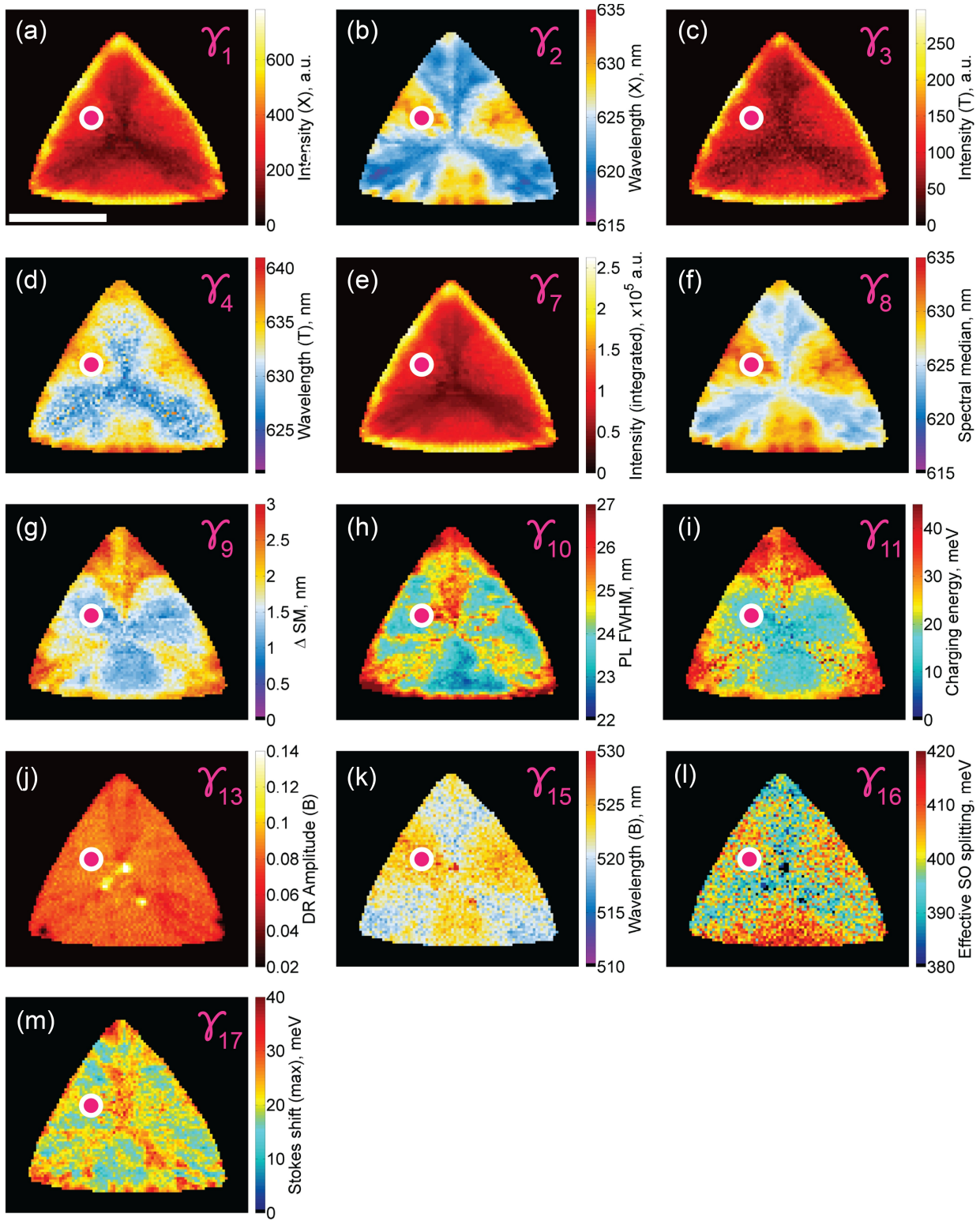

Figure 1: (a–m) Spatial maps corresponding to each of the dimensions of a multidimensional hypercube. A point $\vec{\gamma}$ (purple circle) on the monolayer flake can be represented by a set of 17 numbers, i.e. $\vec{\gamma} = \{\gamma_1, \gamma_2, ..., \gamma_{17}\}$. The length of the scalebar in (a) corresponds to 10 $\mu$m.

4

## S2. Estimation of data densities

To obtain the densities of scattered data-points in each of the 2D projections of a multi-dimensional hypercube, a Gaussian smoothing was used. First, scattered data in a given 2D projection was converted into the $N \times N$ 2D histogram. This ensures that the width of a Gaussian filter will remain the same independent on the nature of the projection under consideration. In other words, each scatter plot was converted into an image $I(x,y)$ of $N \times N$ pixels, thus, standardizing the data analysis between all 2D projections. The value of $I$ corresponds to the number of scattered data-points within a bin. Second, a Gaussian smoothing filter

$$G(x,y) = e^{-\frac{x^2+y^2}{2\sigma^2}} \tag{1}$$

was applied to a $N \times N$ image $I(x,y)$, where $x, y = \{1, 2, ..., N\}$. The smoothing has been performed by the convolution of the image $I(x,y)$ with the Gaussian kernel $G(x,y)$ as follows

$$S(x,y) = \sum_{x'=1}^{N} \sum_{y'=1}^{N} G(x', y') I(x-x', y-y'), \tag{2}$$

where $S(x,y)$ is the smoothed image. The smoothed image was then normalized to the maximum of the data density.

Figure 2 demonstrates the effect of the Gaussian smoothing on the 2D projection formed by the parameters $\gamma_5$ (PL peak intensity) and $\gamma_9$ (PL $\Delta$SM) for different values of $\sigma$ and $N = 400$. In this work, the width of the Gaussian kernel was chosen $\sigma = 12$ unless otherwise specified.

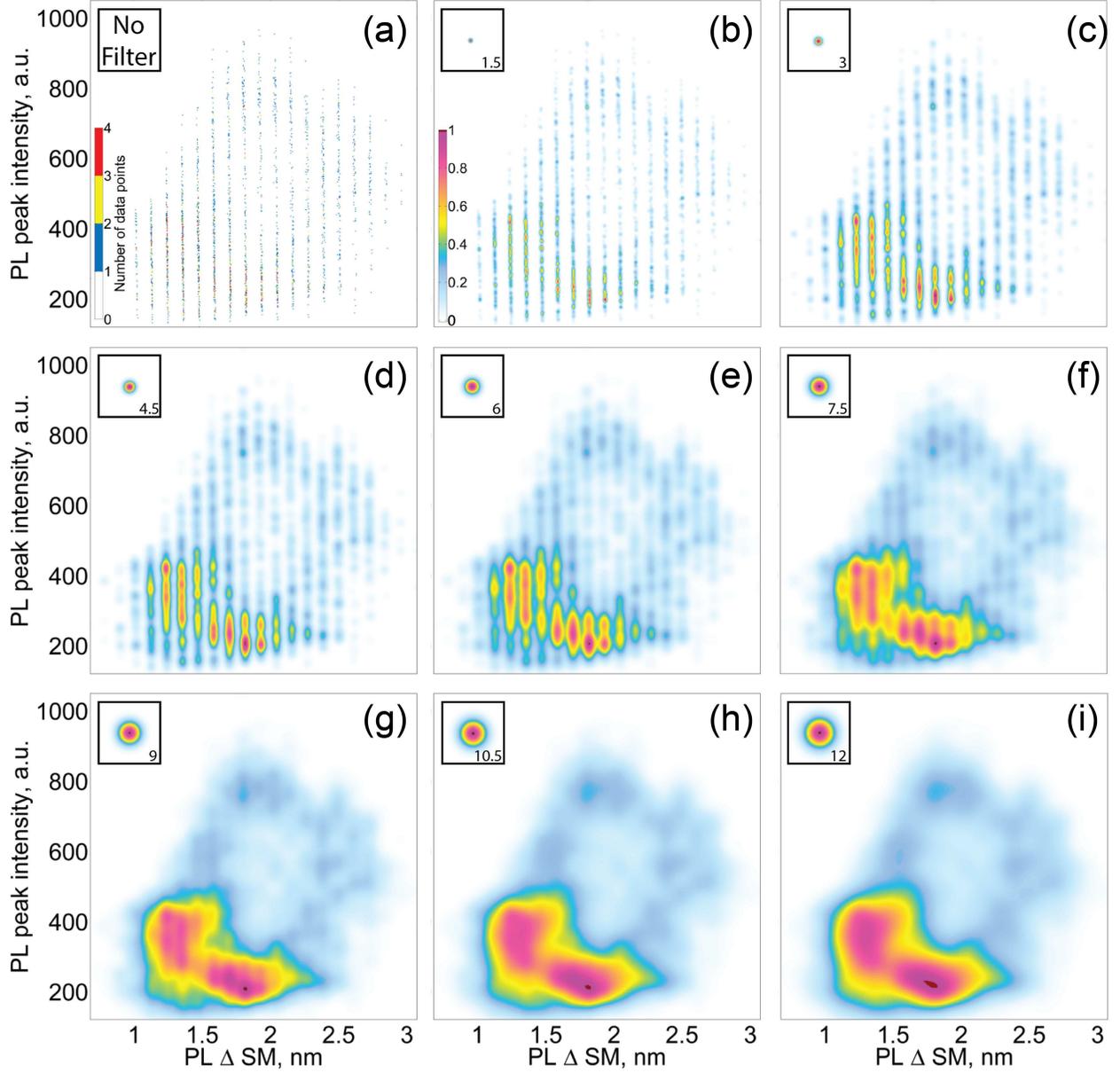

Figure 2: Application of the Gaussian smoothing filter to the projection $(\gamma_5, \gamma_9)$ representing correlation between PL peak intensity and PL $\Delta$SM. (a) Two-dimensional 400x400 histogram I(x,y) of the scattered data with no filter applied. The origin of fringes is the limited spectral resolution of the spectrometer: the distance between the fringes corresponds to the pixels of the spectrometer's detector. The colorbar calibrates the number of data-points within a bin. (b–i) Data density plots for increasing values of $\sigma$. The kernels $G(x,y)$ and their widths $\sigma$ are shown in insets. The colorbar in (b) is applicable to (c–i).

# S3. 2D orthogonal projections featuring structures resembling a torus

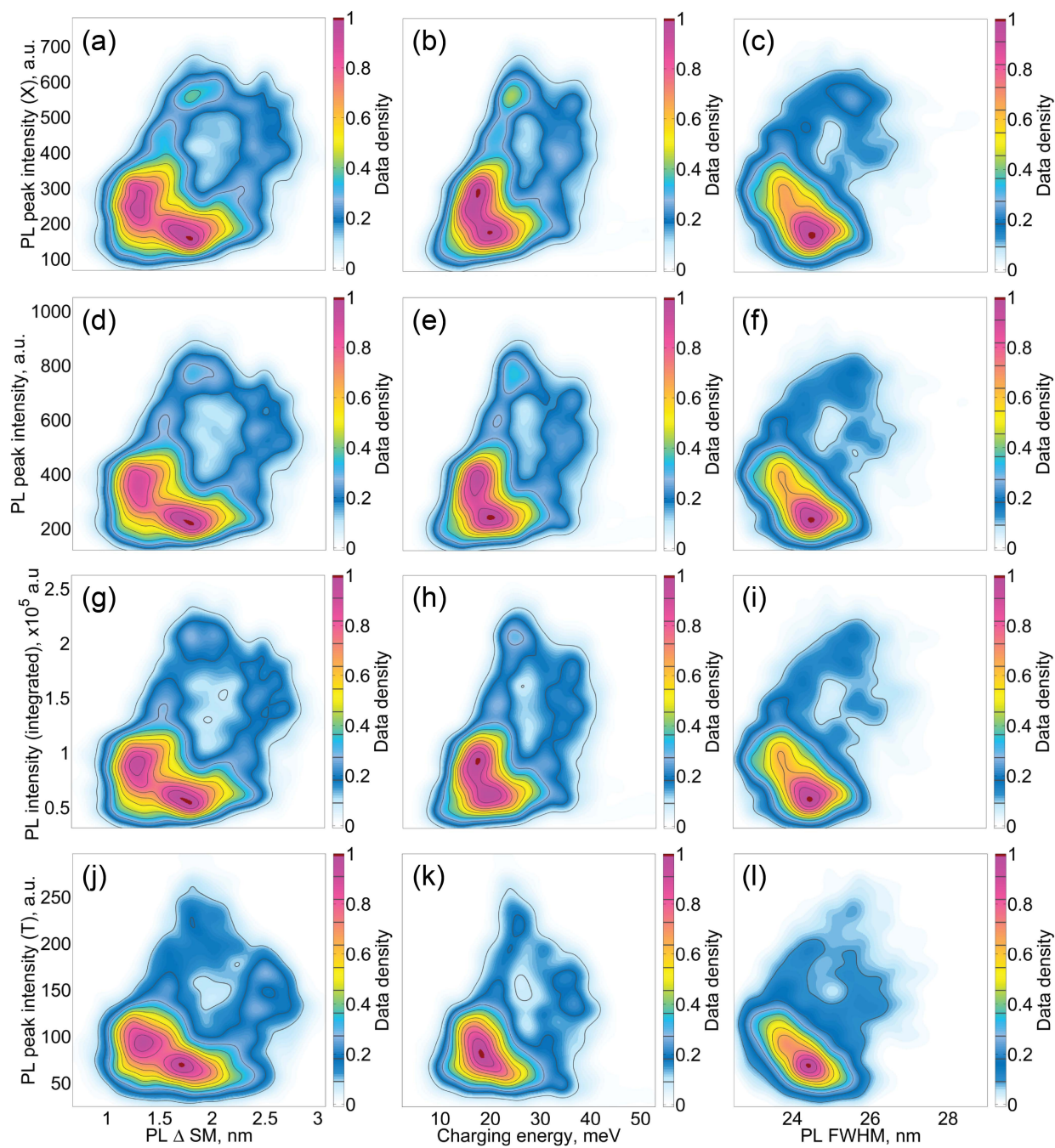

Figure 3: 2D orthogonal projections resembling toroidal shapes of the data-density.

# S4. Negative contour curvatures as signatures of boundaries between the clusters

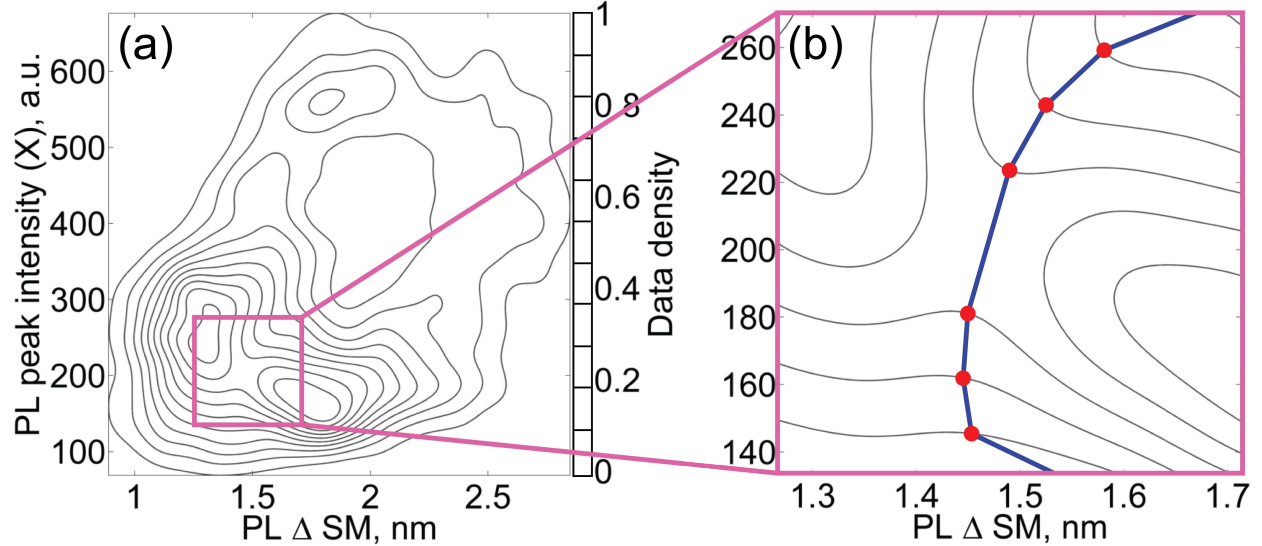

Figure 4: (a) Contour plot of the data-density plot formed from the projection $(\gamma_9, \gamma_1)$. (b) An example of how boundaries between the clusters in the data-cloud could be identified by connecting the point of a high negative curvature.

The data-density plots introduced above could be considered as digital elevation models (DEM) of topographic landscapes[1] (Figure 4a), featuring valleys and channel network. Contours of the data-density plots are closed loops with points having either a negative curvature (concave), a positive curvature (convex) or the zero curvature. The points of a high negative curvature could be a signature of a boundary between the clusters, and connecting such points together (Figure 4b) could render a boundary network. A robust automatic identification of channel network is an ongoing research[1–3] which could be beneficial in the problems of identification of clusters in data-clouds.

# S5. Charging energy as an oblique projection of the data in a hypercube of a lower dimensionality

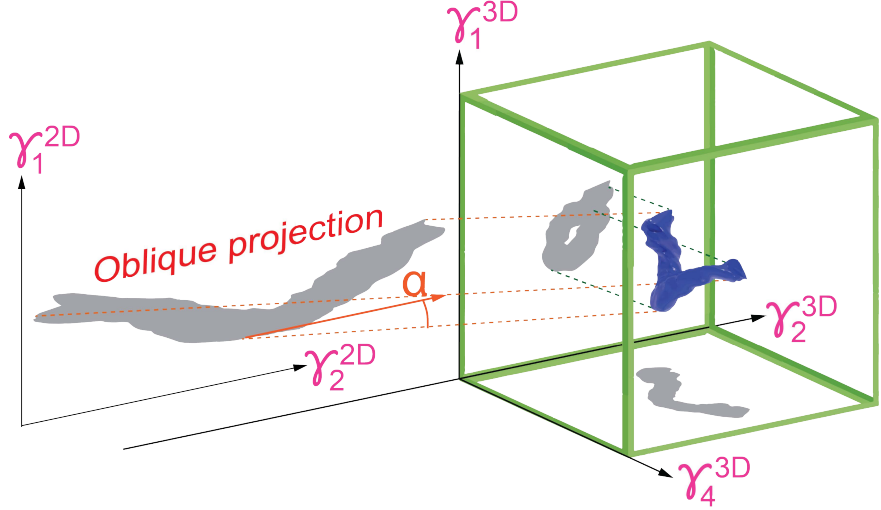

Figure 5: Schematic diagram of a hypercube (green), a data-cloud (blue) and its orthogonal and oblique projections (gray shadows).

Considering the parameters $\gamma_2^{3D}$ (exciton emission energy), $\gamma_4^{3D}$ (trion emission energy) and $\gamma_1^{3D}$ (exciton emission intensity) of a 3-dimensional cube (or an $N$-dimensional hypercube), a linear superposition of $\gamma_2^{3D}$ and $\gamma_4^{3D}$ can be represented as an oblique projection on a 2-dimensional plane (or an $(N-1)$-dimensional hypercube). This follows from the matrix representation of an oblique projection

$$\begin{pmatrix} \gamma_4^{2D} \\ \gamma_2^{2D} \\ \gamma_1^{2D} \end{pmatrix} = \begin{pmatrix} 0 & 0 & 0 \\ -\cot\alpha & 1 & 0 \\ -\cot\beta & 0 & 1 \end{pmatrix} \begin{pmatrix} \gamma_4^{3D} \\ \gamma_2^{3D} \\ \gamma_1^{3D} \end{pmatrix}, \qquad (3)$$

where $\alpha, \beta$ are the angles of obliqueness: $\alpha$ is the angle between the positive direction of $\gamma_2$-axis and the projection lines (dashed orange), projected onto the plane $(\gamma_4, \gamma_2)$ (Figure 5); $\beta$ is the angle between the positive direction of $\gamma_1$-axis and the projection lines, projected onto the plane $(\gamma_4, \gamma_1)$ (in Figure 5, $\beta = \pi/2 \pm \pi n, n \in Z$); $\gamma_4^{3D}, \gamma_2^{3D}, \gamma_1^{3D}$ are coordinates

of the points of a 3D object, and $\gamma_2^{2D}, \gamma_1^{2D}$ are the coordinates of the corresponding points, projected onto the plane $(\gamma_2, \gamma_1)$. The oblique projection given in Eq. (3) can be reduced to the following two equations:

$$\begin{aligned} \gamma_2^{2D} &= \gamma_2^{3D} - \gamma_4^{3D} \cot \alpha, \\ \gamma_1^{2D} &= \gamma_1^{3D} - \gamma_4^{3D} \cot \beta. \end{aligned} \quad (4)$$

If $\alpha = \beta = \pi/2 \pm \pi n, n \in Z$, then the oblique projection becomes degenerate and represents the orthogonal projection of the 3D object onto the plane $(\gamma_2, \gamma_1)$. One can notice that in cases when $\alpha = \pm \pi/4 \pm \pi n, n \in Z$, and $\beta = \pi/2 \pm \pi n, n \in Z$, Eqs. (4) further reduce to

$$\begin{aligned} \gamma_2^{2D} &= \gamma_2^{3D} \mp \gamma_4^{3D}, \\ \gamma_1^{2D} &= \gamma_1^{3D}. \end{aligned} \quad (5)$$

The first expression in Eqs. (5), in the case of "−" sign, is a definition of the trion charging energy CE ($\gamma_2^{2D}$) and therefore this parameter can be considered as a result of an oblique projection. Charging energy, therefore, adopts the resultant non-trivial stretching effect and makes trends and clusters easier to observe.

# S6. Principal component analysis

Principle component analysis (PCA) has been described elsewhere.[4–6] Shortly, the procedure was reduced to the following steps.

1. First, the $m \times N$ matrix $\tilde{X}$ representing $m$ data-points in an $N$-dimensional hypercube was normalised so that the normalised dataset $X$ had zero mean. To be more specific, each element $\tilde{x}_{i,j} \in \tilde{X}, i = \{1, ..., m\}, j = \{1, ..., N\}$ was transformed into the element $x_{i,j}$ as follows:
$$x_{i,j} = \frac{\tilde{x}_{i,j} - \tilde{\mu}_j}{\tilde{\sigma}_j}, \tag{6}$$
where $\tilde{\mu}_j$ and $\tilde{\sigma}_j$ are the mean and the standard deviation of the data along the dimension $j$, respectively.

2. Second, the normalised dataset $X$ was presented as the product of three matrices via the singular value decomposition (SVD) method[7] as
$$X = USU, \tag{7}$$
where $U$ is the $N \times N$ matrix of eigenvectors $u_i$ (principal components (PC)), and $S$ is the diagonal $N \times N$ matrix of eigenvalues $s_{ii}$, $i = \{1, ..., N\}$. The columns $u_i$ of the matrix $U$ define unit vectors of a new $N$-dimensional hypercube which is a rotated version of the initial hypercube, and, therefore, are linear combinations of unit-vectors $\gamma_i, i = \{1, ..., N\}$ (see Table 2–3 and Figure 7). The eigenvalues $s_{ii}$ describe the variance of the data along the axis $u_i$ (see Table 4).

3. Finally, the data $X$ has been projected onto the 2D plane defined by the first two principal components along which the variance of the data is the largest compared to that along the rest of the principle components (Figure 6):
$$Z = XU_2, \tag{8}$$

where $Z$ is the projection of the $N$-dimensional data $X$ onto the 2D plane spanned by the unit-vectors $u_1$ and $u_2$, and $U_2$ is $N \times 2$ matrix of these eigenvectors.

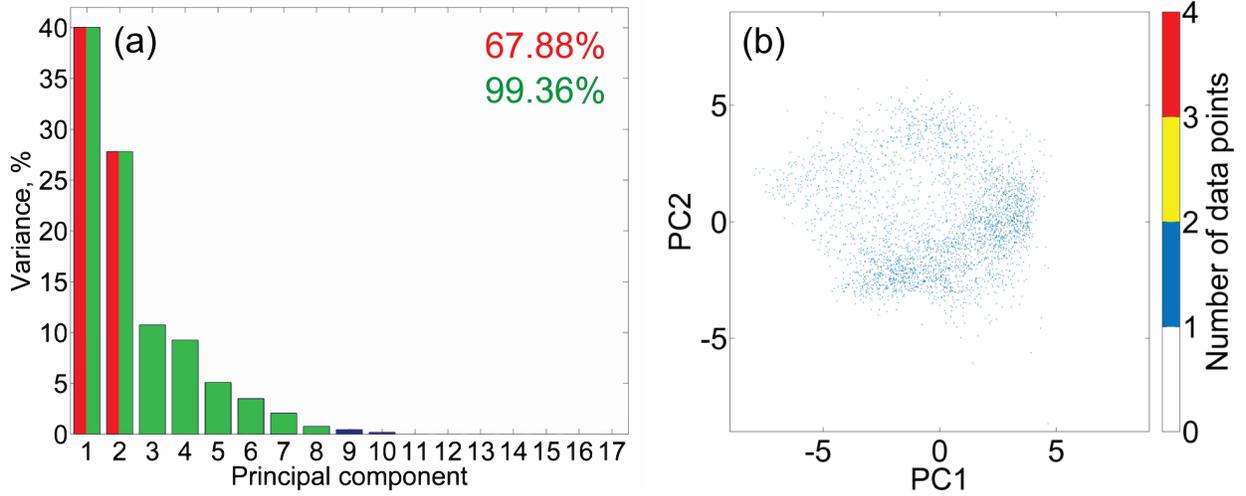

Figure 6: (a) Variances of the data along each of the principal components. 99.36% of the overall variance is retained in 8-dimensional PC-hypercube (green); 67.88% of the overall variance is retained in 2D PC-plane (red) used for visualization of data. (b) 2D histogram representing the data in the plane spanned by the first two principle components.

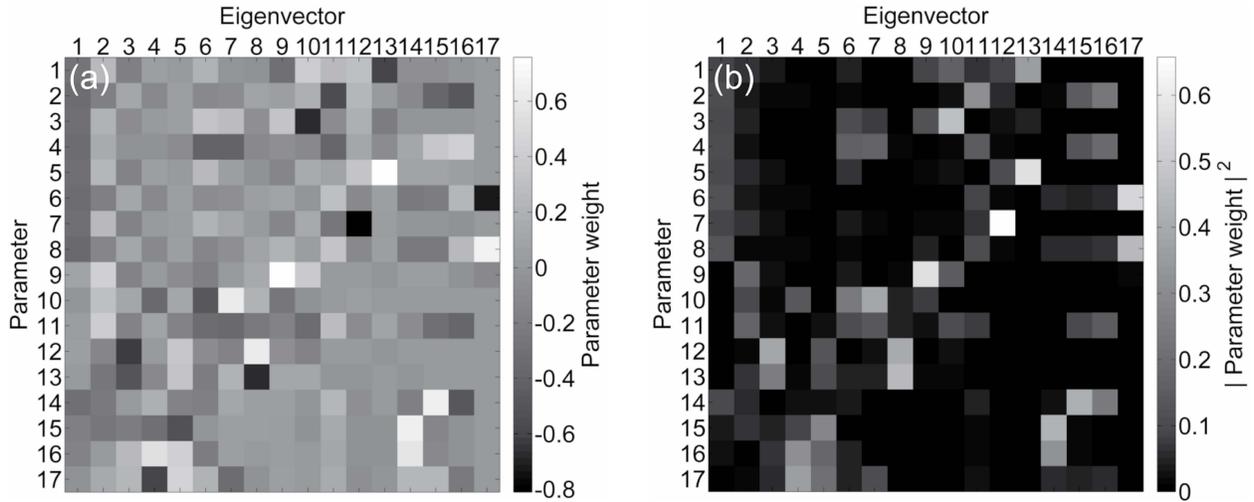

Figure 7: Graphical representation of the eigenvectors $u_i$ determined from PCA. (a) Each eigenvector is a linear combination of unit vectors $\gamma_j$ of the initial hypercube axes with coefficients $c_{i,j}$ (the weights of the parameters $\gamma_j$) color-coded in the shades of gray. (b) Squared values of parameter weights.

We note that we have not excluded multilayers from our PCA-analysis due to two reasons: (i) the number of corresponding data-points is much smaller compared to the overall number

12

Table 2: Eigenvectors ($u_1 - u_9$) found by singular value decomposition of the 17-dimensional data-cloud. The numbers are coefficients $c_{i,j}$ of a linear superposition $u_i = \sum_{j=1}^{N} c_{i,j} \gamma_j$.

|  | $u_1$ | $u_2$ | $u_3$ | $u_4$ | $u_5$ | $u_6$ | $u_7$ | $u_8$ | $u_9$ |
|---|---|---|---|---|---|---|---|---|---|
| $\gamma_1$ | -0.2941 | 0.2496 | -0.1963 | 0.0696 | -0.0005 | 0.2089 | -0.0188 | -0.0302 | -0.3029 |
| $\gamma_2$ | -0.3349 | -0.1978 | 0.1051 | -0.1026 | 0.0618 | -0.1032 | -0.0828 | 0.0871 | 0.0562 |
| $\gamma_3$ | -0.3139 | 0.2035 | -0.0875 | 0.0216 | 0.0582 | 0.3269 | 0.2815 | -0.0559 | 0.3383 |
| $\gamma_4$ | -0.3258 | 0.1491 | -0.0300 | -0.0477 | -0.0788 | -0.4028 | -0.4178 | -0.1040 | -0.0753 |
| $\gamma_5$ | -0.3077 | 0.2325 | -0.1703 | 0.0582 | 0.0182 | 0.2500 | 0.0575 | -0.0263 | -0.1430 |
| $\gamma_6$ | -0.3377 | -0.1888 | 0.1191 | -0.1068 | 0.0614 | -0.0971 | -0.0529 | 0.0476 | 0.0838 |
| $\gamma_7$ | -0.3028 | 0.2520 | -0.1547 | 0.0213 | 0.0258 | 0.1988 | 0.1063 | 0.0028 | -0.1404 |
| $\gamma_8$ | -0.3540 | -0.1358 | 0.1040 | -0.1147 | 0.0510 | -0.1356 | -0.0566 | 0.0754 | 0.2183 |
| $\gamma_9$ | 0.0872 | 0.4183 | -0.1544 | 0.0110 | -0.0954 | -0.1765 | 0.0067 | 0.1398 | 0.7552 |
| $\gamma_{10}$ | -0.0043 | 0.3129 | 0.1192 | -0.3680 | 0.0965 | -0.4881 | 0.6196 | 0.2037 | -0.2707 |
| $\gamma_{11}$ | 0.0450 | 0.4071 | -0.1616 | 0.0727 | -0.1643 | -0.3253 | -0.3677 | -0.2232 | -0.1508 |
| $\gamma_{12}$ | 0.0659 | -0.1400 | -0.6187 | 0.0111 | 0.3560 | -0.0786 | -0.1687 | 0.6345 | -0.0677 |
| $\gamma_{13}$ | 0.0111 | -0.2580 | -0.5042 | -0.1197 | 0.3326 | -0.2069 | 0.2131 | -0.6668 | 0.1043 |
| $\gamma_{14}$ | -0.3074 | -0.2369 | 0.0074 | 0.1731 | -0.1498 | -0.1839 | 0.0988 | 0.0666 | 0.0491 |
| $\gamma_{15}$ | -0.1777 | -0.2571 | -0.2126 | -0.3018 | -0.5310 | -0.0078 | 0.0577 | 0.0631 | 0.0241 |
| $\gamma_{16}$ | -0.1730 | 0.0044 | 0.2524 | 0.5592 | 0.4269 | -0.2170 | 0.0573 | 0.0054 | 0.0302 |
| $\gamma_{17}$ | -0.0267 | 0.1275 | 0.2319 | -0.6038 | 0.4581 | 0.2040 | -0.3302 | -0.0494 | 0.0671 |

Table 3: Eigenvectors ($u_{10} - u_{17}$) determined from singular value decomposition of the 17-dimensional data-cloud. The numbers are coefficients $c_{i,j}$ of a linear superposition $u_i = \sum_{j=1}^{N} c_{i,j} \gamma_j$.

|  | $u_{10}$ | $u_{11}$ | $u_{12}$ | $u_{13}$ | $u_{14}$ | $u_{15}$ | $u_{16}$ | $u_{17}$ |
|---|---|---|---|---|---|---|---|---|
| $\gamma_1$ | 0.4008 | 0.2574 | 0.2928 | -0.5924 | -0.0684 | -0.0516 | -0.0155 | 0 |
| $\gamma_2$ | 0.1709 | -0.5604 | 0.2335 | 0.0154 | -0.1046 | -0.3906 | -0.4832 | 0 |
| $\gamma_3$ | -0.6812 | -0.0957 | 0.1699 | -0.2054 | -0.0194 | -0.0147 | -0.0090 | 0 |
| $\gamma_4$ | -0.1099 | -0.3835 | 0.1142 | -0.0944 | 0.0979 | 0.3582 | 0.4358 | 0 |
| $\gamma_5$ | 0.1496 | 0.0824 | 0.3334 | 0.7585 | 0.0894 | 0.0744 | 0.0495 | 0 |
| $\gamma_6$ | -0.0236 | 0.3006 | -0.1065 | 0.0598 | -0.2310 | -0.2185 | 0.2360 | -0.7326 |
| $\gamma_7$ | 0.1353 | -0.2676 | -0.8110 | 0.0252 | -0.0047 | -0.0103 | -0.0281 | 0 |
| $\gamma_8$ | 0.0384 | 0.3296 | -0.1141 | 0.0638 | -0.2452 | -0.2321 | 0.2510 | 0.6712 |
| $\gamma_9$ | 0.3816 | 0.0089 | 0.0126 | -0.0090 | 0.0415 | 0.0383 | -0.0396 | -0.1128 |
| $\gamma_{10}$ | -0.0372 | 0.0109 | 0.0688 | 0.0007 | 0.0006 | 0.0014 | 0.0008 | 0 |
| $\gamma_{11}$ | -0.3295 | 0.2715 | -0.0889 | 0.0957 | -0.0846 | -0.3179 | -0.3908 | 0 |
| $\gamma_{12}$ | -0.1559 | 0.0164 | 0.0026 | -0.0030 | 0.0011 | -0.0001 | -0.0008 | 0 |
| $\gamma_{13}$ | 0.1020 | -0.0089 | -0.0039 | 0.0029 | -0.0022 | 0.0003 | 0.0006 | 0 |
| $\gamma_{14}$ | -0.0177 | 0.2242 | -0.0680 | 0.0026 | -0.1752 | 0.6486 | -0.4898 | 0 |
| $\gamma_{15}$ | -0.0270 | 0.1370 | -0.0481 | -0.0497 | 0.6562 | -0.1409 | -0.0372 | 0 |
| $\gamma_{16}$ | 0.0090 | 0.1190 | -0.0446 | -0.0432 | 0.5709 | -0.1226 | -0.0324 | 0 |
| $\gamma_{17}$ | 0.0011 | 0.1688 | -0.0531 | -0.0236 | 0.2385 | 0.2255 | -0.2403 | 0 |

Table 4: Eigenvalues $s_{ii}$ determined from singular value decomposition of the 17-dimensional data-cloud.

| $s_{1,1}$ | 6.8075 | $s_{5,5}$ | 0.8670 | $s_{9,9}$ | 0.0763 | $s_{13,13}$ | $8 \cdot 10^{-5}$ | $s_{17,17}$ | $6 \cdot 10^{-17}$ |
|---|---|---|---|---|---|---|---|---|---|
| $s_{2,2}$ | 4.7278 | $s_{6,6}$ | 0.5968 | $s_{10,10}$ | 0.0305 | $s_{14,14}$ | $2 \cdot 10^{-5}$ | | |
| $s_{3,3}$ | 1.8281 | $s_{7,7}$ | 0.3550 | $s_{11,11}$ | 0.0010 | $s_{15,15}$ | $9 \cdot 10^{-6}$ | | |
| $s_{4,4}$ | 1.5733 | $s_{8,8}$ | 0.1307 | $s_{12,12}$ | 0.0006 | $s_{16,16}$ | $6 \cdot 10^{-6}$ | | |

of data-points, so that the presence of multilayer-related data-points influence eigenvectors and eigenvalues negligibly, and (*ii*) we wished to capture multilayers in the PC1-PC2 plane for the sake of generality.

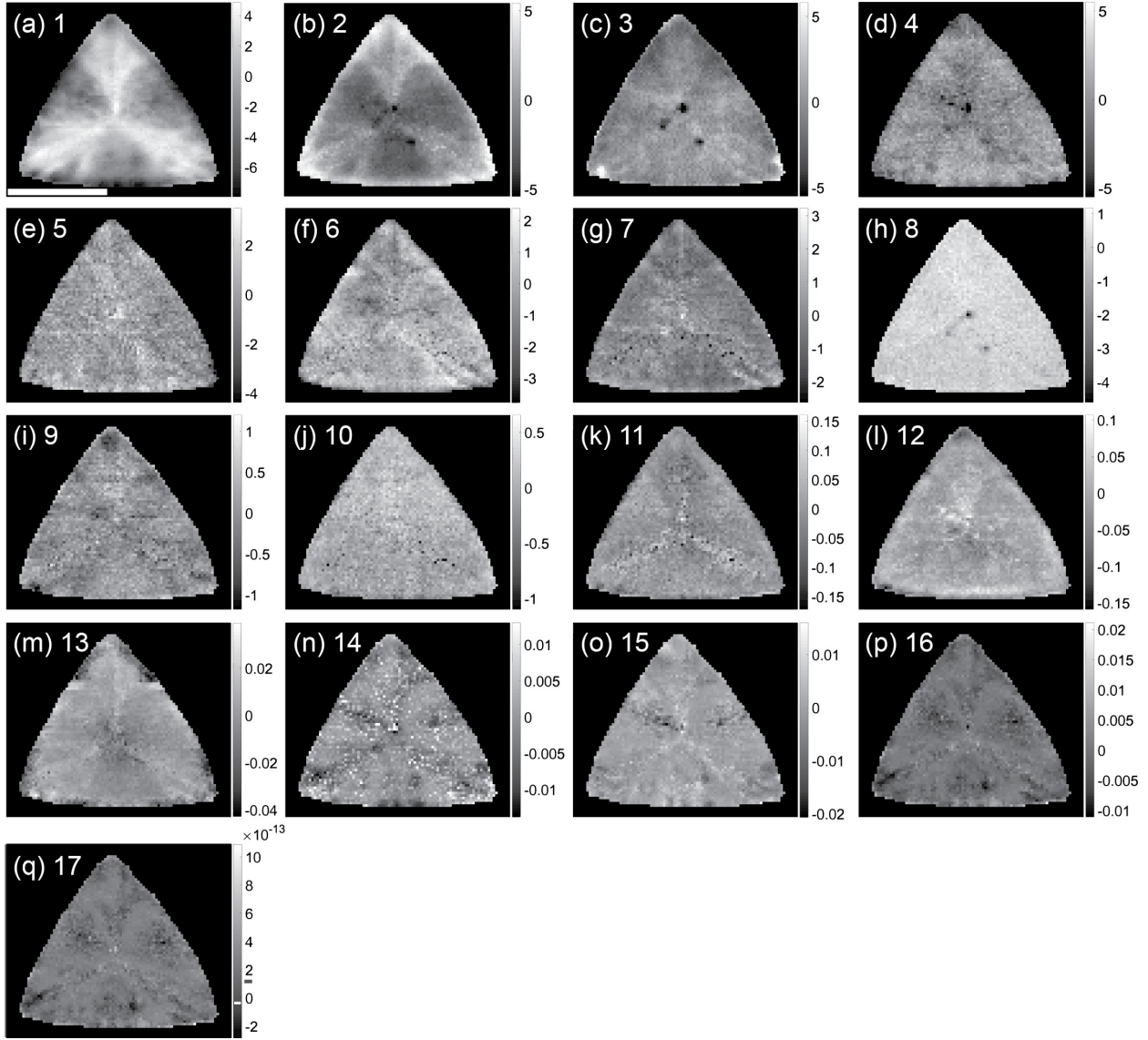

Figure 8: (a–q) Spatial maps of principal components. The length of the scalebar in (a) corresponds to 10 $\mu$m.

## S7. Identification of multilayers as anomalies (outliers)

The orthogonal projections of the data on the sides of the $N$-cube with the $y$-axis formed by either A- or B-exciton absorption peak amplitudes feature data-points lying outside of and above the main body of the data (high data-density regions). Two of such projections are shown in Figure 9. The data-points above the high data-density body correspond to the regions on the monolayer flake with large absorption amplitude and reflect the dependence of the absorption peak amplitudes on the number of layers reported previously.[8–12] Therefore, we assign these domains and corresponding data-points to multilayer $WS_2$ material since the absorption peak amplitudes change discontinuously within those regions. It is worth noting that multilayer data-points in Figure 9 feature a wavelength-dependent behaviour having larger wavelengths in the bright interior regions and lower wavelength in the dark regions which is likely due to coupling effects between adjacent layers.

In this work, to identify the multilayer domains we considered A-B-exciton amplitude-amplitude correlation plot (Figure 10) since both amplitudes increase with the number of layers. Within unsupervised learning approach multilayers were considered as anomalies in a method similar to the anomaly detection methods described elsewhere.[13] The core idea of an anomaly detection algorithm is to fit the data with a probability distribution function and label those data-points as anomalies that correspond to low values (below a specified threshold) of this function. Here, instead of fitting the data, we apply a Gaussian smoothing filter (Eq. (2)) with $\sigma = 12$ to the projection leaving the data-cloud featureless and resembling a multivariate Gaussian distribution. The result then was normalised and a threshold of 0.065 was chosen to detect "anomalous" data-points. This detection was performed in a directional way as shown in Figure 10: only those points were considered as anomalies which lie within the yellow-shaded area. This area was chosen to be bound by two semi-infinite lines starting at the maximum of the data-cloud with each line passing through the point of a negative curvature of the outermost contour.

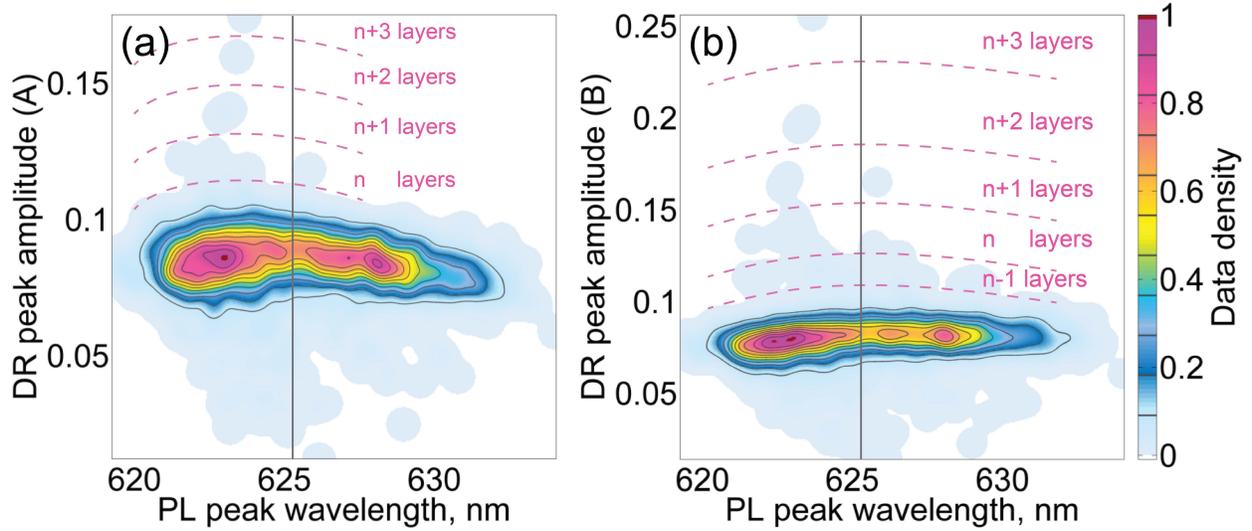

Figure 9: 2D projections of a hypercube formed by (a) A- and (b) B-exciton absorption peak amplitudes and PL peak wavelength. The vertical gray line splits the data-points in two parts with one part (left) originating from dark regions on the monolayer flake and another part originating from the bright regions. Dashed pink lines are guidelines separating different multilayer clusters.

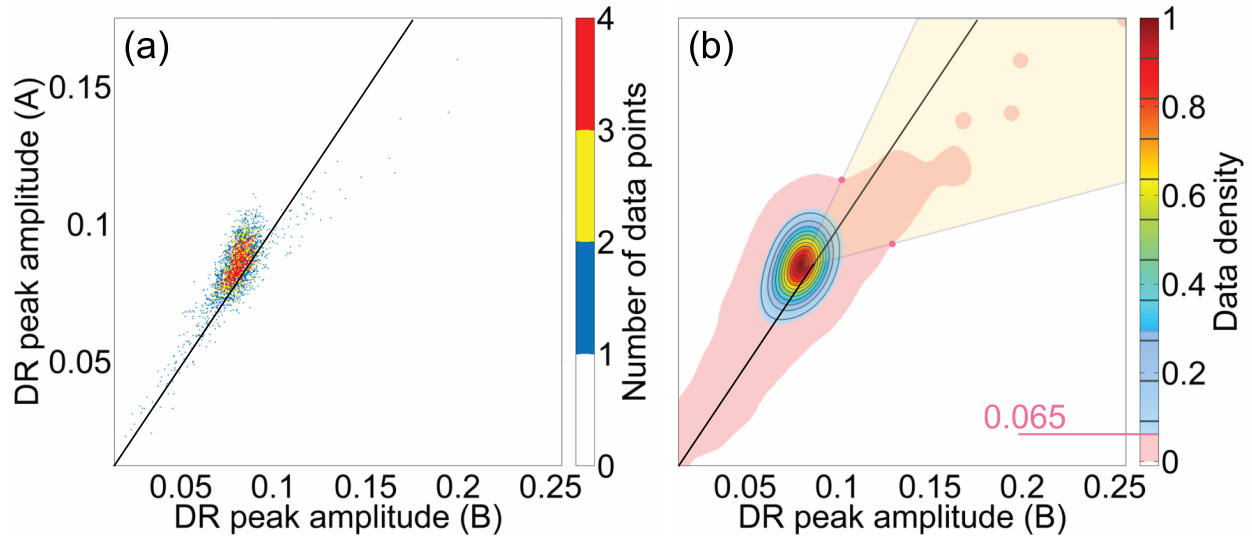

Figure 10: (a) 2D histogram showing correlation between A- and B-exciton absorption peak amplitudes. (b) 2D histogram after application of the Gaussian smoothing filter with $\sigma = 7$. The result was normalised, and the values below the threshold of 0.065 were colorcoded in pink. The yellow area corresponds to the area where multilayers were searched in the anomaly detection method. Two pink points correspond to the points of negative curvatures of the contours defining the silhouette of the data-cloud. Black lines in (a–b) represent diagonals.

# S8. Determination of the number of clusters for K-means clustering

It is well-known that K-means algorithm requires the value of $K$ as an input defined by a user having some *a priori* knowledge/insight about the nature of the data-cloud, and it is not always straightforward to identify $K$. In this work, two methods have been considered in efforts to identify the number of clusters present in the PCA-projection of the data shown in Figure 6b.

**(i) Elbow method**

Elbow method is, perhaps, the most popular method for determination of optimal number of clusters in a K-means clustering method and was described elsewhere.[14] Shortly, for an input number of clusters $K$, the K-means clustering algorithm iteratively tries to minimise the following objective function $J$:

$$J(C_1, ..., C_m, \mu_1, ..., \mu_K) = \sum_{i=1}^{m}[(x_{i,1} - \mu_{C_i,1})^2 + (x_{i,2} - \mu_{C_i,2})^2], \qquad (9)$$

where $K$ is the number of clusters; $C_i = \{1, ..., K\}$ is the index of the cluster in the PCA-plane to which the data-point $x_i = \{x_{i,1}, x_{i,2}\}$ belongs to; $\mu_k = \{\mu_{k,1}, \mu_{k,2}\}, k = \{1, ..., K\}$ is the point in the PCA-plane representing the centroid of the cluster $k$; $\mu_{C_i} = \{\mu_{C_i,1}, \mu_{C_i,2}\}$ is the point in the PCA-plane representing the centroid of the cluster $C_i$ to which the data-point $x_i$ has been assigned; $m$ is the number of data-points.

Elbow method is the method of inspection of the dependence $J(K)$ of the minimised objective function $J$ on the number of clusters $K$. If for a given dataset a natural set of clusters exists then the graph $J(K)$ will feature an obvious "elbow" (Figure 11a) formed by the change of slope of the function $J(K)$. In cases when there exist several natural sets of clusters then there may be several "elbows" present in the plot $J(K)$. In the case of WS$_2$ monolayer considered in this work, there are at least two elbows present in the plot $J(K)$,

as evidenced in the logarithmic plot of the function $J(K)$ shown in Figure 11b.

Each point in Figure 11a,b was calculated as follows. To make sure that $J$ corresponds to the global minimum (as opposed to a local minimum), for a given $K$, 5000 random initialisations of the positions of $K$ centroids were performed, and for each of these initialisations the function $J$ was calculated. Amongst all $J$-functions for the given $K$ the minimum value was found and plotted as a point in the graph $J(K)$. This procedure was repeated for all $K = \{1, ..., 25\}$.

To identify the optimal numbers of clusters $K$ more precisely and in an automatic manner we used the following procedure.

1. For each segment $[K, K+1], K = \{1, ..., 24\}$, calculate the slope of the segment $[\log_{10}(J(K)), \log_{10}(J(K+1))]$ (Figure 11c).

2. Calculate the elbow strengths as differences between the slopes at $K$ and $K-1$ for all $K = \{2, ..., 24\}$. Assign the elbow strength to zero at $K = 1$ (Figure 11d).

3. Compute the squares of elbow strengths and find local maxima above a certain threshold (0.0004 in our case) in the resulting plot (Figure 12).

This procedure allowed us to identify 4 natural cluster sets (purple points in Figure 12) that can be used to fit the data-cloud in the PCA-plane using K-means clustering: these sets correspond to $K = 2, 4, 8, 12$ with the most natural set consisting of 4 clusters ($K = 4$).

### (ii) Method of local maxima in the data density landscape

For a naturally-identifiable cluster set, it is also possible to find the value of $K$ by smoothing the scattered data with a Gaussian kernel of a certain width and finding the local maxima of the resultant landscape (Figure 13b,d,f,h). The procedure of finding the local maxima was the following. For a constant threshold of the data density (0.1 in this work) the width $\sigma$ of the Gaussian smoothing filter (Eqs. 1–2) was kept increasing by an increment of 0.1 until the desired number of local maxima was identified. Starting from $\sigma = 7.0$, the case of $K = 12$

occurred at $\sigma = 9.7$, the case of $K = 8$ occurred at $\sigma = 11.7$, the case of $K = 4$ occurred at $\sigma = 17$, and the case of $K = 2$ occurred at $\sigma = 33$. In Figure 13, the comparison between the positions of the centroids that minimise the objective function $J$ (Figure 13a,c,e,g) and the positions of local maxima of the data-density landscape (Figure 13b,d,f,h) is shown. Compared to the "elbow" method, the method of local maxima in the data-density plot allows to initialise centroids for a given $K$ in a straightforward and intuitive manner without performing multiple random centroid initialisations. In Figure 11a,b, red points correspond to the values of $J$ where centroids were initialised at these local maxima in the K-means calculations showing that the difference is of the order of $10^{-5} - 10^{-4}$.

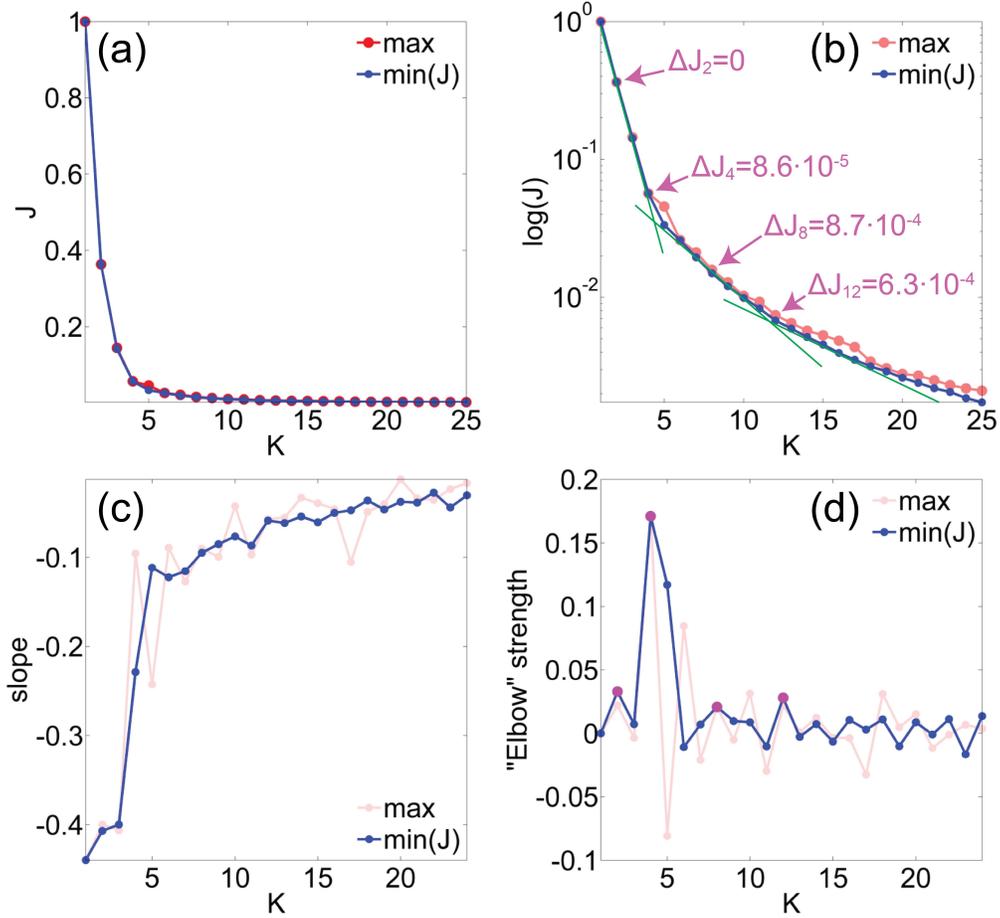

Figure 11: (a) Dependence $J(K)$ of the objective function $J$ on the number of centroids $K$. (b) Dependence $J(K)$ represented on a logarithmic scale. (c) Slopes of each of the line segments in (b) versus the number of centroids. (d) Elbow strengths versus the number of centroids $K$. In (a,b) red points correspond to the case where prior to K-means clustering centroids were initialised at the local maxima of the data-density plots (see the local maxima method below). The differences $\Delta J$ between the two objective functions evaluated for $K = 2, 4, 8, 12$ were found to be $0, 8.6 \cdot 10^{-5}, 8.7 \cdot 10^{-4}$ and $6.3 \cdot 10^{-4}$, respectively. In (b), three colored lines serve as guidelines showing different slopes of the function $\log_{10}(J(K))$. Purple points in (c–d) mark the number of clusters identified from calculations of elbow strengths; the red dim points correspond to the case where centroids were initialized at the local maxima of the data-density plot.

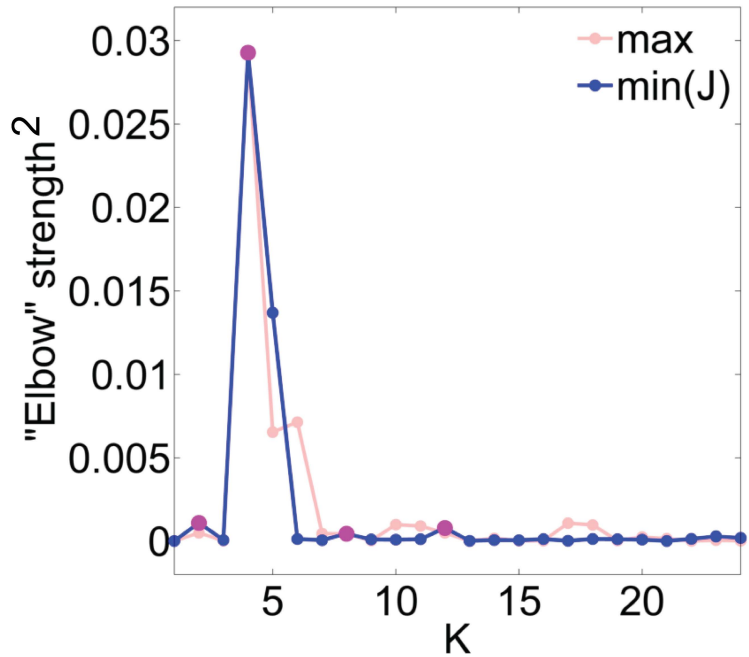

Figure 12: Dependence of the squared values of elbow strengths on the number of clusters. Purple points mark the local maxima at $K = 2, 4, 8, 12$ located above the threshold of 0.0004. The dependence of the squared "elbow" strength on the number of clusters initiated at the local maxima of the data-density plot is shown in dim red. This dependence was normalized to the maximum of the blue curve.

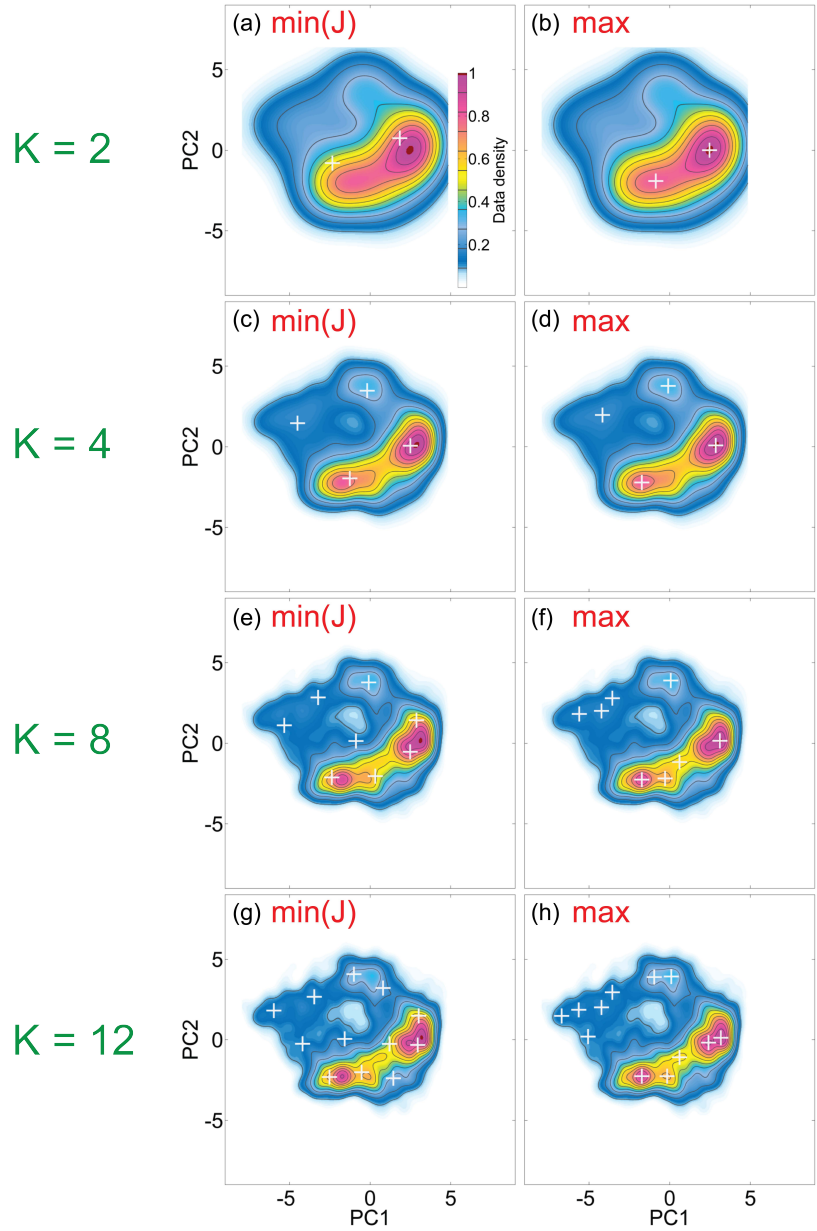

Figure 13: Comparison between the positions of centroids leading to the global minimum of the objective function $J$ (left column: a,c,e,g) and the initial positions of centroids identified from the local maxima method (right column: b,d,f,h). The width of the smoothing filters in the cases of $K = 2, 4, 8, 12$ were $\sigma = 33, 17, 11.7, 9.7$ for $K = 2, 4, 8, 12$, respectively.

# S9. K-means algorithm and the results for $K = 2, 4, 8, 12$

K-means clustering algorithm has been described elsewhere.[15–17] Simply, it consists of the following steps.

1. For an initialised set of centroids $\mu_k = \{\mu_{k,1}, \mu_{k,2}\}, k = \{1, ..., K\}$, split all data-points into $K$ different clusters in accordance with whether a data-point is closer to the centroid $\mu_1, \mu_2, ..., \mu_{K-1}$ or $\mu_K$.

2. Compute the averages between the data-points assigned to same clusters, and use these averages as new centroid positions.

3. Repeat the steps 1 and 2 until the centroids converge to a minimum of the objective function $J$ (Eq. (9)).

The number of iterations (step 3) in this work was chosen to be 30.

We excluded multilayer-related data points from the K-means clustering analysis, because in this case these data-points affect the result significantly.

Figure 14 shows the clustering results for $K = 2, 4, 8, 12$ in the cases when the function $J(K)$ is at its absolute minimum (Figure 14a,b,e,f,i,j,m,n) and when centroids were initialised at the local maxima of the data-density plot (Figure 14c,d,g,h,k,l,o,p). It is seen that for the case of $K = 2$, two main domains corresponding to the dark and bright regions on the monolayer flake have been clearly separated. A fine structure is introduced in the case of $K = 4$ clusters revealing the distinct and heterogeneous edge. In the case of $K = 8$, more "shades" of optoelectronic properties are introduced revealing a domain (blue) that lies between the four main domains in the phase and real spaces. Further refinement can be observed for the case of $K = 12$.

Both centroid initialisation methods described above produced identical clustering results for $K = 2$, and nearly identical results for $K = 4$. For $K = 8$, however, centroids initialised by the local maxima method (Figure 14k,l) did not lead to the identification of the central

blue domain as in the case of the absolutely minimised function $J(K)$ (Figure 14i,j); instead, a shade of green at the apexes has been introduced. In the case of the absolute minimum of $J(K)$, this shade is introduced only for $K = 12$ (Figure 14m,n). For $K = 12$, the differences between the two approaches of centroid initialisation is the most obvious (Figure 14m–p), however, qualitatively, the domains have been successfully identified in both cases.

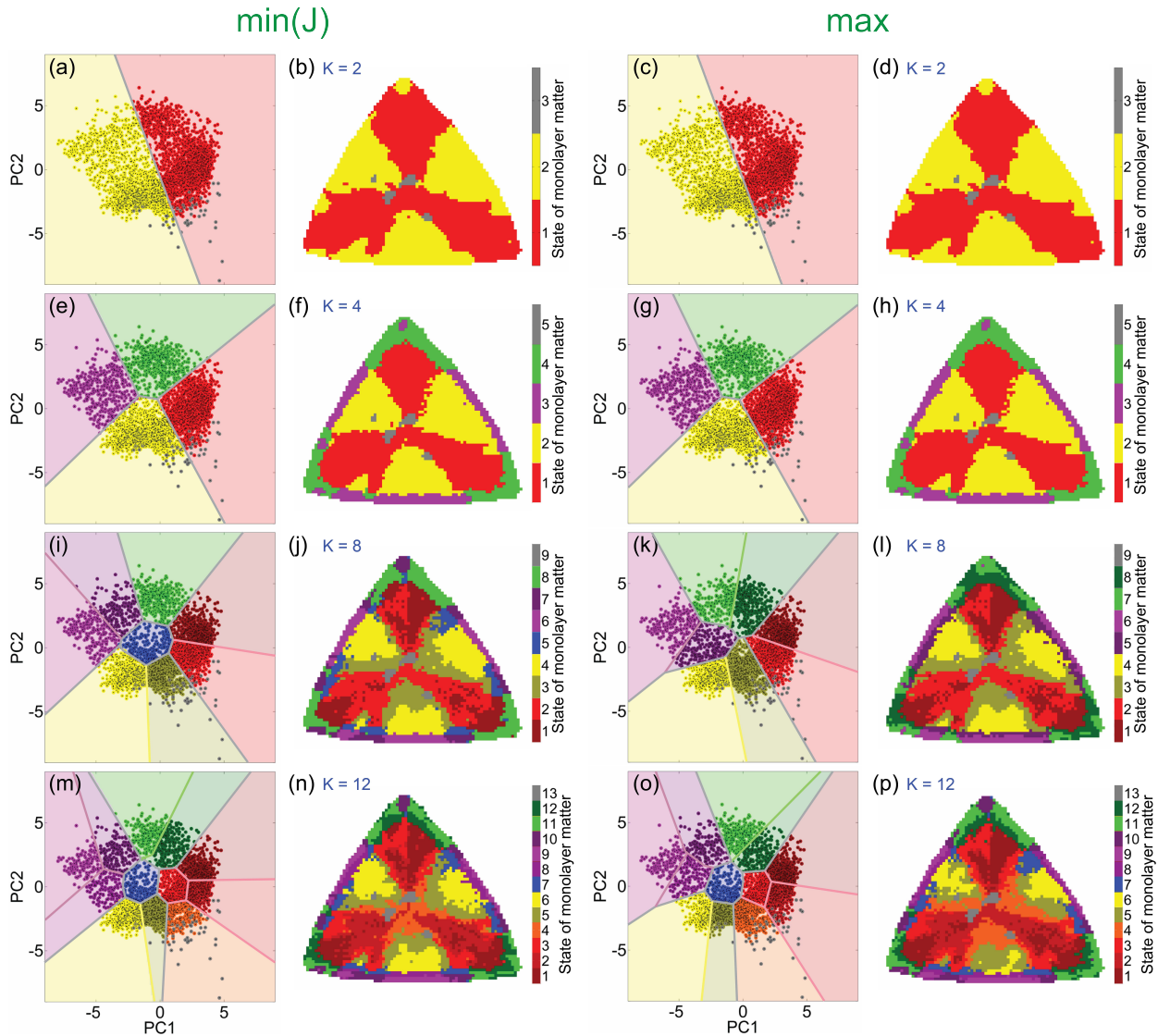

Figure 14: K-means clustering results: a comparison between the two methods of cluster initialisation. Two columns (a–m) and (b–n) on the left correspond to the case of the initial cluster locations that correspond to the global minimum of the function $J$. The two columns (c–o) and (d–p) on the right shows the clustering results when centroids were initialised at the local maxima of the data density plots. The cases of (a–d) $K = 2$, (e–h) $K = 4$, (i–l) $K = 8$ and (m–p) $K = 12$ are shown.

# S10. Evidence that water intercalation has not released strain

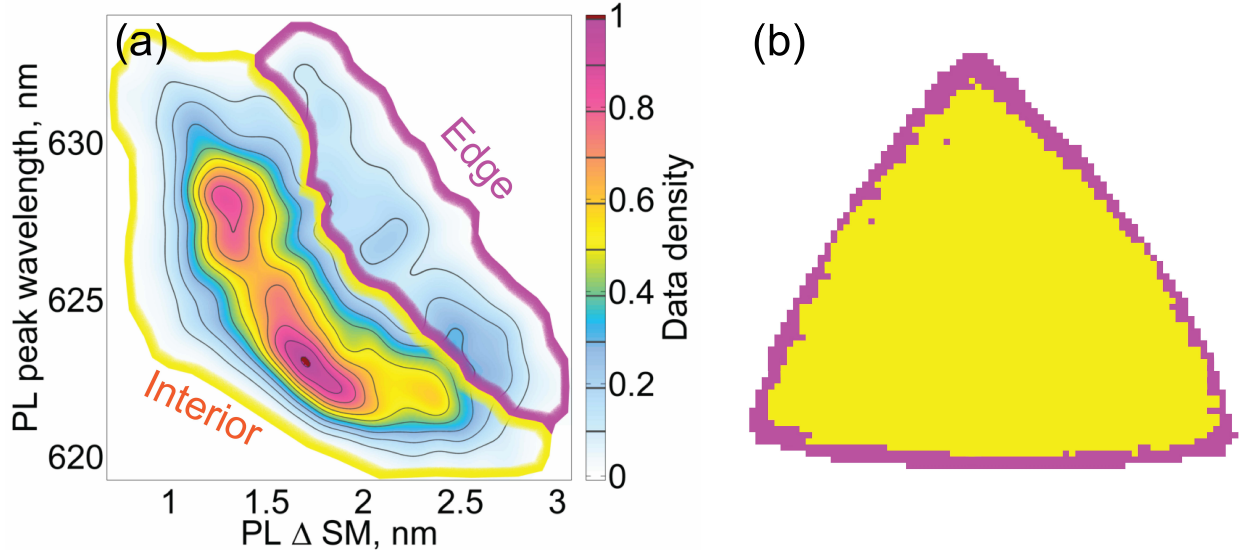

Figure 15: Orthogonal projection of the multi-dimensional data-cloud onto the plane spanned by the parameters $\gamma_6$ (PL peak wavelength) and $\gamma_9$ ($\Delta$SM). Two regions has been drawn approximately following trenches in the data-density landscape separating high- and low-density areas. (b) Mapping of the data-points back onto the monolayer flake demonstrates that water intercalation has not released intrinsic strain present in the crystal structure.

# S11. Other approaches capable of introducing dimensions to a hypercube

Table 5: Other approaches capable of introducing dimensions to a hypercube.

| # | Method | Dimension(s) | Refs. |
|---|---|---|---|
| 1 | Raman imaging | Parameters of vibrational modes (intensity, frequency) | [18] |
| 2 | Polarization-resolved PL[a] | Degree of the circular polarization | [19] |
| 3 | SNOM[b] PL[a]/absorption | PL[a]/absorption spectral parameters | [20–22] |
| 4 | Electric-field assisted SNOM[b] PL[a] | Bias voltage | [23] |
| 5 | Opto-valleytronic imaging | Valley polarization, valley coherence | [24] |
| 6 | micro-PLE[c] spectroscopy | PLE[c] spectral parameters | [25] |
| 7 | PSHG/SHG[d] imaging and fitting models | Orientation of an armchair direction, crystal orientation, strain field parameters (amplitude and direction) | [26–28] |
| 8 | CARS[e] microscopy | CARS[e] intensity | [29,30] |
| 9 | FWM[f] microspectroscopy | Exciton radiative/dephasing lifetimes, degree of the circular polarization, doping level | [31] |
| 10 | SF-2DES[g] | Characteristics of 0Q[h], 1Q[h], 2Q[h] spectra (e.g. coupling strength between quantum states) | [32] |
| 11 | Nanoscale ARPES[i] | Parameters of electronic dispersion $E(k)$[j] (bandwidth, effective mass, band alignment) | [33,34] |
| 12 | KPFM[k] | Electronic surface potential | [34,35] |
| 13 | AFM[l] | Height above a substrate AFM[l] phase | [35] |
| 14 | Nanoscale XPS[m] | Defect density | [34] |
| 15 | Temperature-dependent | Length of the trion spectral tail | [36] |
| 16 | trion fitting model | due to electron recoil effects | |
| 17 | TEM[n] and GPA[o] | Strain field parameters (orientation and amplitude) | [37] |
| 18 | TR-PEEM[p] | Carrier decay time constants | [38] |

[a] PL = Photoluminescence; [b] SNOM = Scanning near-field optical microscopy; [c] PLE = Photoluminescence excitation; [d] (P)SHG = (Polarization-resolved) second harmonic generation; [e] CARS = Coherent anti-Stokes Raman scattering; [f] FWM = Four-wave mixing; [g] SF-2DES = Spatially-resolved fluorescence-detected two-dimensional electronic spectroscopy; [h] 0Q = Zero-quantum; 1Q = One-quantum; 2Q = Two-quantum; [i] ARPES = Angle-resolved photoemission spectroscopy; [j] $E$ is the binding energy; $k$ is the wavevector; [k] KPFM = Kelvin probe force microscopy; [l] AFM = Atomic force microscopy; [m] XPS = X-ray photoelectron spectroscopy; [n] TEM = Transmission electron microscopy; [o] GPA = Geometrical phase analysis; [p] TR-PEEM = Time-resolved photoemission electron microscopy.



\end{document}